\documentclass[conference]{vgtc}         

\usepackage{authblk}
\ifpdf
  \pdfoutput=1\relax                   
  \usepackage{graphicx}                
  \DeclareGraphicsExtensions{.pdf,.png,.jpg,.jpeg} 
\else
  \usepackage{graphicx}                
  \DeclareGraphicsExtensions{.eps}     
\fi%

\graphicspath{{figures/}{pictures/}{images/}{./}} 
\usepackage{multirow}
\usepackage{makecell}
\usepackage{colortbl}
\usepackage{hhline}
\usepackage{verbatimbox}
\usepackage{microtype}                 
\PassOptionsToPackage{warn}{textcomp}  
\usepackage{textcomp}                  
\usepackage{mathptmx}                  
\usepackage{times}                     
\usepackage{cite}                      
\usepackage{tabu}                      
\usepackage{booktabs} 

\usepackage{amsmath}
\usepackage{multirow} 

\usepackage{hyperref}  
\hypersetup{pdfborder={0 0 0} }  

\usepackage{enumitem}
\usepackage{amsmath}
\usepackage{setspace}

\usepackage{color}
\onlineid{1185}

\vgtccategory{Application}

\vgtcpapertype{Design Study}

\title{SilkViser: A Visual Explorer of Blockchain-based Cryptocurrency Transaction Data\vspace{-0.5cm}}

\author[1,2]{Zengsheng Zhong}
\author[1]{Shuirun Wei}
\author[1]{Yeting Xu}
\author[1]{Ying Zhao}
\author[1]{Fangfang Zhou}
\author[1]{Feng Luo}
\author[1]{Ronghua Shi\thanks{Email: zzs@ctbu.edu.cn, \{0904130202, 0902160219, zhaoying, zff, luofeng365, shirh\}@csu.edu.cn. Ying Zhao is the corresponding author. }}

\affil[1]{\normalsize{School of Computer Sciences and Engineering, Central South University, China;}}
\affil[2]{\normalsize{Chongqing Engineering Technology Research Center for Information Management in Development, \authorcr Chongqing Technology and Business University, China}\vspace{-0.5cm}}

\abstract{Many blockchain-based cryptocurrencies provide users with online blockchain explorers for viewing online transaction data. However, traditional blockchain explorers mostly present transaction information in textual and tabular forms. Such forms make understanding cryptocurrency transaction mechanisms difficult for novice users (NUsers). They are also insufficiently informative for experienced users (EUsers) to recognize advanced transaction information. This study introduces a new online cryptocurrency transaction data viewing tool called SilkViser. Guided by detailed scenario and requirement analyses, we create a series of appreciating visualization designs, such as paper ledger-inspired block and blockchain visualizations and ancient copper coin-inspired transaction visualizations, to help users understand cryptocurrency transaction mechanisms and recognize advanced transaction information. We also provide a set of lightweight interactions to facilitate easy and free data exploration. Moreover, a controlled user study is conducted to quantitatively evaluate the usability and effectiveness of SilkViser. Results indicate that SilkViser can satisfy the requirements of NUsers and EUsers. Our visualization designs can compensate for the inexperience of NUsers in data viewing and attract potential users to participate in cryptocurrency transactions. } 


\begin{document}
\firstsection{Introduction}
\maketitle
Exponential growth of the cryptocurrency community has been witnessed in recent years since the first real-world Bitcoin transaction occurred in 2010 \cite{A1}. Novel types of cryptocurrencies are being continuously created. Millions of people are rushing to participate in cryptocurrency transactions and to hold on cryptocurrency assets. As of June 2019, the number of different cryptocurrencies available on the Internet has reached over 1,900 \cite{A2}. Bitcoin, as the world's first and largest cryptocurrency, has more than 7 million active users. Moreover, 60\% of Americans have heard of Bitcoin and approximately 5\% own Bitcoin assets \cite{A3}.

Cryptocurrencies are commonly built upon variations of blockchain technology. A blockchain acts as a public decentralized ledger for validating and recording transactions that ever happened on the blockchain, and the generated transaction data is publicly accessible to everyone. In 2010, Bitcoin released the first open-source web tool, called Blockchain Explorer \cite{A4}, to help people view their Bitcoin transaction data online. This tool has become an informal standard in the cryptocurrency community. Using Blockchain Explorer as a reference, numerous cryptocurrencies have launched their blockchain explorers, such as Zcash Blockchain Explorer \cite{A5} and Litecoin Block Explorer \cite{A6}, to provide online transaction data viewing services.

Most blockchain explorers present information about blocks, transactions, and addresses in transaction data in textual and tabular forms. However, the shortcomings of only using such presentation forms have emerged from two major aspects: unfriendliness for novice users (NUsers) and insufficient informativeness for experienced users (EUsers). The transaction systems of cryptocurrencies completely differ from those of fiat currencies. For NUsers who are interested in cryptocurrencies but lack experience in cryptocurrency transactions, viewing transaction data through blockchain explorers are a natural step toward understanding necessary concepts (e.g., blockchain, block, and six-confirmation) and processes (e.g., mining, coin mixing, and chain-forming) related to operating mechanisms of cryptocurrency transaction systems. However, the textual and tabular presentations of transaction data are minimally helpful in deepening their understanding for an easy beginning of participating in cryptocurrency transactions. EUsers who are skilled in operating explorers (i.e., clients, miners, and managers), propose new requirements for recognizing advanced information in daily trading. For example, users frequently inquire which inputs or outputs are important in a transaction, miners are highly interested in the fee distribution of transactions in a block, and managers should know the recent generating trends of blocks and transactions. However, such information cannot be effectively presented in textual and tabular forms.

This study proposes to introduce visualization techniques into traditional blockchain explorers to overcome the two aforementioned shortcomings. Adopting the practice of the cryptocurrency Silubium \cite{A7} in upgrading its blockchain explorer, namely, SilkViewer \cite{A7}, as background, we design a new blockchain explorer, called SilkViser \cite{A7}, for the online viewing of the Silubium transaction data. SilkViser consists of four levels of web pages (i.e., blockchain, block, transaction, and address pages). Each web page uses a set of appreciating visualizations to help users intuitively understand relevant concepts and processes and visually recognize essential and advanced information about four types of objects (i.e., blockchain, block, transaction, and address objects) abstracted from the Silubium transaction data. For example, we adopt the metaphor of a string of paper ledgers to design a blockchain visualization and a block glyph in the blockchain page to explain concepts related to the blockchain and blocks and to illustrate the chain-forming and block-confirming processes, we use a copper coin-inspired glyph design to encode four essential information fields of a transaction in the block page, we design a coin-Sankey diagram in the transaction page to explain the concepts of coin mixing and highlight important inputs and outputs in a transaction. Moreover, we provide a set of lightweight interactions in SilkViser to support easy and free data viewing, mainly including brushing and sorting interactions as well as two web page exploring modes (i.e., hierarchical and query exploring modes).

The entire design processing of SilkViser is conducted in a user-centered way and considers adaptability to other cryptocurrencies. For the user-centered design, we initially conducted a series of in-depth pilot communications with a wide range of target users, including Silubium cofounders, managers, clients, and potential clients, to understand scenarios, abstract data, formulate requirements, and conclude design principles. Then, we worked alongside with several target users to iteratively evolve our visualization and interaction designs. Lastly, we conducted a controlled user study to quantitatively evaluate the usability and effectiveness of SilkViser. The results indicate that SilkViser can satisfy the requirements of NUsers and EUsers. For adaptability to other cryptocurrencies, our scenario analysis and data abstraction focus on data characteristics and user requirements common in diverse cryptocurrencies. Concepts, processes, and advanced information that should be presented through visualizations can also be found in other cryptocurrency transaction data. These efforts ensure that SilkViser can be used to explore the transaction data of other cryptocurrencies with minor modifications and that our visualization design can be shared among a variety of blockchain explorers.

In summary, this study introduces a new online blockchain explorer called SilkViser. A user-centered manner is adopted in this study to formulate user requirements, and navigate the design and evaluation process. A series of novel visualizations are created in SilkViser to help users understand cryptocurrency transaction mechanisms and recognize advanced transaction information.

%

\vskip 0.3cm
\section{Related Work}
\subsection{Blockchain Data Analysis}
Cryptocurrency transaction data is a major type of blockchain data, which has elicited increasing concerns among a wide range of researchers due to their accessibility. Several researchers have attempted to evaluate and compare core characteristics of cryptocurrencies, such as decentralization \cite{B1}, anonymity \cite{B2,D3,D6}, and security \cite{D1,D2}, through transaction data analysis. For example, Gencer et al. \cite{B3} evaluated the decentralization level of Bitcoin and Ethereum by adapting decentralization metrics to transaction data. M\"oser et al. \cite{B4} compared the anonymity of three different Bitcoin coin-mixing services based on Bitcoin transaction data analysis. Some researchers have become interested in user transaction behavior patterns. For example, Ron and Shamir \cite{B5}analyzed typical user behavior patterns of acquiring and spending Bitcoins. Meiklejohn et al. \cite{B6} introduced heuristic clustering algorithms to identify transaction addresses that may belong to the same user. Security is another popular research topic in blockchain data analysis. For example, Vasek and Moore \cite{D1} investigated financial fraud in Bitcoin transactions. Lim et al. \cite{D2} pointed out the risks that cryptocurrency transaction systems suffer from cyberattacks.

An exchange is an organized market wherein tradable cryptocurrencies are sold and bought with effective safety regulations. The analysis of exchange data provides people with a comprehensive understanding of trading market developments and timely warnings of transaction risks \cite{B12}. For example, Kiran et al. \cite{B7} and Gronwald et al. \cite{B8} analyzed and featured the price fluctuations of Bitcoin exchanges through exchange data analysis. Athey et al. \cite{B10} developed a data-driven model for predicting the variation tendency of Bitcoin price. Researchers have also analyzed other types of blockchain data, such as social media data collected from online cryptocurrency communities \cite{B13} and questionnaire data surveyed from Bitcoin users \cite{B14}.

\subsection{Blockchain Data Visualization}
Visualization has become an increasingly important research area due to its wide range of applications in many disciplines \cite{F1,F2,F3,F4,F5,F8,F9,F10,F11,F12,F13,F14,F15,F16,F17,F18}. Recently, the visualization and visual analytics community is eliciting considerable interests in blockchain data presentation and analysis, leading to several pioneering studies \cite{F19}. We introduce them from the perspectives of using statistical diagrams, information visualizations, and visual analytics.

Several researchers conducted statistical analysis on blockchain data and adopted statistical diagrams to explain their results. Elbahrawy et al. \cite{C20} used statistical diagrams to present the market share distribution and the turnover of important cryptocurrencies since April 2013 based on exchange data analysis. Lischke and Fabian \cite{C6} used statistical diagrams to depict evolution patterns of the Bitcoin economy in business categories, such as gambling, donation, and vendors. Bitnodes \cite{C7} and Wizbit \cite{C8} are tools that can present the popularity of Bitcoin worldwide in the forms of 2D and 3D world maps, respectively. In addition, a number of websites, which combine transaction and exchange data, have launched visual dashboards \cite{C10,C11,C13,C14} that show the asset rankings and price fluctuations of various cryptocurrencies.

Some enthusiasts of blockchain and visualization technologies have proposed interesting information visualizations and animations to demonstrate the generation process of Bitcoin blocks and transactions. For example, Interaqt \cite{C3} introduced an emerging bubble chart to present the generation of new transactions, in this chart, bubble color is deduced from transaction hash and bubble size indicates transaction size. Bitbonkers \cite{C4} introduced a dropping pearl animation. When a new transaction or a new block is generated, a colored ball or cube is lively dropped on a fixed plate, providing users with a vivid feast. BitcoinCity \cite{C5} presented a novel city metaphor in which the primary chain of a cryptocurrency is represented by an urban avenue, and transactions or blocks are represented by various buildings along the avenue.

Current visual analysis approaches mainly focus on exploring user transaction behaviors and analyzing transaction anomalies. Battista et al. \cite{C15} designed a tailored flowchart to analyze Bitcoin transactions and illustrate suspicious Bitcoin capital flows. McGinn et al. \cite{C16} proposed a force-directed graph visualization to explore unexpected high-frequency transaction patterns in the Bitcoin transaction network. Pham and Lee\cite{C17} created collaborative Bitcoin transaction graphs to detect illegal and fraudulent transactions and related suspects. Chawathe and Sudarshan \cite{C18} introduced a visual tool based on self-organizing maps to facilitate the detection of fraudulent activities and the diagnosis of anonymity and traceability of Bitcoin transactions. BlockChainVis \cite{C19} is a multiview visual analytic tool that allows users to filter and select the desired part of Bitcoin transactions and then interactively explore correlation patterns of different roles in Bitcoin transactions. Yue et al. \cite{C21} provided a mature and complete visual analysis system, called BitExTract, which can analyze and compare the evolutionary transaction patterns of Bitcoin exchanges among exchange markets. Xia et al. \cite{C22} presented an interactive visual analytics system that supports surveillance of the mining pool and de-anonymization by visual reasoning. Isenberg et al. \cite{D4,D5} developed a series of interactive visual interfaces for analyzing transaction histories of individual entities and financial activities on the Bitcoin transaction network. 

The difference between the aforementioned studies and our work can be summed up in two aspects. First, the previous approaches of visual analysis and statistical analysis are geared toward data analysts, but our work is designed for general users who might not have an IT background. Second, the previous information visualization approaches illustrate only one or two relevant concepts and are independent of blockchain explorers, but our work integrates visual illustrations of concepts and processes related to the blockchain, blocks, transactions, and addresses into a blockchain explorer.
\section{Scenario and Data Abstraction}
\subsection{Motivating scenario}
Silubium \cite{A7} is a currency that is exchanged only in digital form and maintained by a peer-to-peer network with open membership. Silubium is technically derived from Bitcoin; thus, it retains Bitcoin's basic characteristics. Meanwhile, it widely adopts the essential features of other cryptocurrencies, such as the smart contract function in Ethereum. Its monetary unit is notated by SLU, and the primary chain is called Silkchain.

Silubium provides an HTML5 web tool, called SilkViewer \cite{A7}, to enable its clients to view the Silubium transaction data. SilkViewer's design is mostly referenced from Blockchain Explorer \cite{A4}, the most popular open-source online viewing tool for the Bitcoin transaction data. SilkViewer's interface lists the 10 latest transactions and 10 blocks in real-time and allows clients to search for a historical block or transaction by inputting keywords. All real-time and historical information is presented in textual and tabular forms. 

However, SilkViewer's shortcomings have emerged from two major aspects: unfriendliness for NUsers and insufficient informativeness for EUsers. A revision plan was recently launched by Silubium's founders and managers. They aimed to update the back-end data process functions of SilkViewer and integrate visualization techniques into its front-end interface to satisfy the newly emerged user requirements. We were invited to participate in the revision plan, responsible for data abstraction, requirement analysis, visualization design, and evaluation.
\subsection{Data Abstraction}\label{data abstraction}
\begin{figure*}[htb]
	\centering
	\vspace{-0.05cm}  
	\setlength{\belowcaptionskip}{-0.4cm}   
	\includegraphics[width=\textwidth]{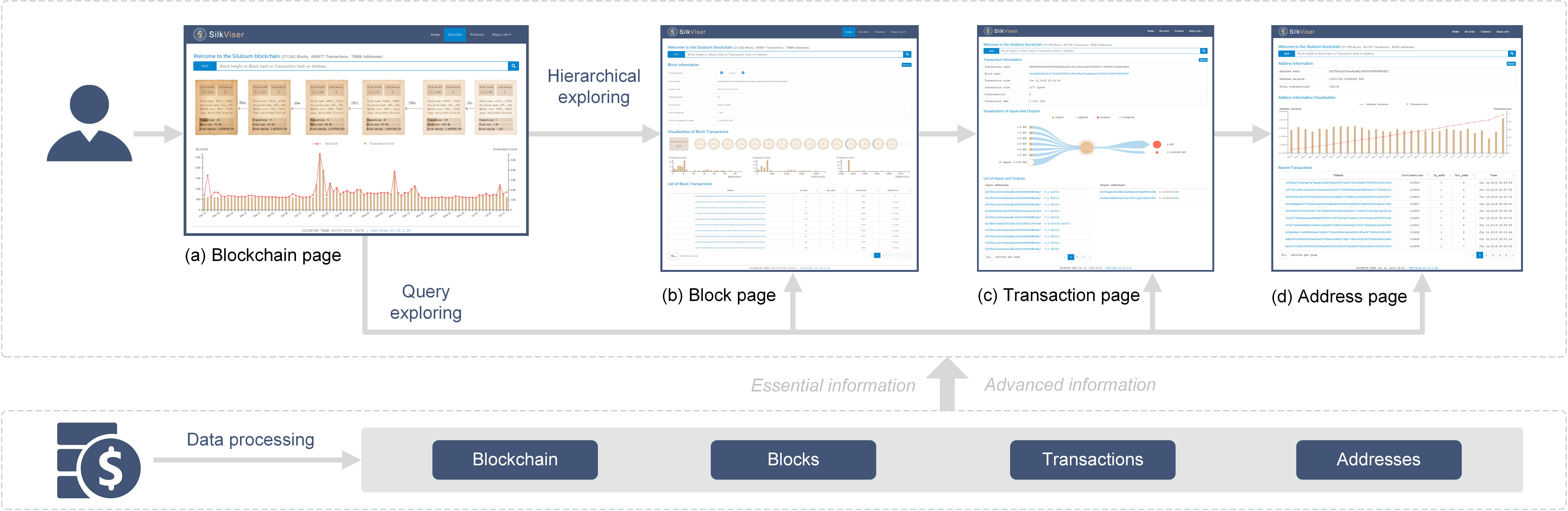}
	\caption{Overview of SilkViser. It consists of two parts: data processing and user interface. The data processing extracts the essential and advanced information of the four types of objects, namely, blockchain, block, transaction, and address, from the Silubium transaction data. The user interface provides four levels of web pages (a--d) corresponding to the four types of objects respectively and supports two kinds of web page exploring modes, namely query and hierarchical exploring. }
	\label{figure1}
\end{figure*}
The Silubium transaction data is a kind of multi-object, multi-attribute tabular data that holds all transaction information from the beginning until the present. In general, the data contains four types of objects: blockchain, block, transaction, and address. Each object has several or even dozens of basic information fields. The four objects form a four-level hierarchical structure. In particular, a blockchain is a chain of linked blocks. A block generally records many transactions, and a transaction may involve multiple participating addresses. The concepts associated with and the forming processes related to the four objects are keys to understanding the Silubium transaction system. Most cryptocurrencies, including Bitcoin and Ethereum, also share these concepts and processes. We briefly introduce the important concepts and processes as follows.

(1) Address: An address is a cryptographic identifier presented by a 36-character encrypted string and regarded as an individual Silubium account. A Silubium client commonly possesses more than one account that can be used as inputs to send SLU or outputs to receive SLU.

(2) Transaction and coin mixing: A transaction presented by a 64-character encrypted string is inputs and outputs, and each input or output is associated with an SLU amount. When a transaction involves multiple inputs, the inputting SLU is mixed by the Silubium transaction system to unlink input and output addresses. Such a coin-mixing service improves the anonymity of cryptocurrency transactions.

(3) Block: A block is a digital account book that records a collection of transactions issued within a certain time interval (2 min by default in Silubium) and occupies a certain amount of storage space (maximum 8 \textit{M} by default in Silubium). A newly generated block is broadcast through the network and validated by each network node. 

(4) Chain and chain-forming: A blockchain is a sequence of blocks that is gradually formed by each new block linked to the previous block (further reinforcing the validity of all previous blocks), except for the genesis block. A blockchain functions as a public ledger because it maintains the integrity of all the transaction records and a copy of which can be maintained at each network node.

(5) Confirming and six-confirmation: Confirming is a process in which a block is continuously validated. Once a new block is validated by the entire network, it gains its first confirmation. When a subsequent block is linked to this block, the confirmation number of this block is added. When the confirmation number reaches six or more, this block and all the transactions it contains are computationally infeasible to modify. 

(6) Mining and mining reward: Mining refers to the process in which network nodes, called miners, solve increasingly difficult computation problems to validate new transactions and pack them into a new block. Nodes that complete successful mining are rewarded. Block reward comes from two sources. The first source is the transaction fees from all the transactions in the new block. Second, the system automatically mints a certain amount of block subsidy (newly available digital currency) for each new block generation.

\section{User Requirements and Design Principles}
\subsection{Design Requirements}
We interviewed seven individuals, including two domain experts, two EUsers, and three NUsers, to understand how users use SilkViewer and identify opportunities to improve SilkViewer. We identified the following three major design requirements of SilkViser based on the feedback gathered in the interviews.

\textbf{DR1:} Essential information refinement and advanced information extraction. The abstraction of the Silubium transaction data has obtained four types of objects with numerous information fields. Empirically, many fields are trivial for users' daily data viewing. NUsers and EUsers expect SilkViser to mainly display essential information fields. Meanwhile, EUsers want to obtain advanced information (e.g., trends and distributions) of the four types of objects from the data.

\textbf{DR2:} Visual illustrations of important concepts and processes. NUsers hope that SilkViser could adopt appropriate visual metaphors and design intuitive visual glyphs to encode the essential information of the four types of objects. These visual information encodings can help them understand important concepts and processes for an easy beginning in participating in Silubium transactions.

\textbf{DR3:} Visual presentations of advanced information. The advanced information that EUsers focus on is mostly about various trends and distributions related to the four types of objects. Such information is difficult to effectively convey using traditional textual and tabular forms. EUsers hope that SilkViser could adopt appropriate visualization diagrams to present advanced information to help them accurately and efficiently recognize such information.

\subsection{Design Principles}
We also summarized three design principles that we should adopt in SilkViser design. The principles are as follows.

\textbf{DP1:} Considering the independence and relevance of data objects. The four types of objects, namely, blockchain, block, transaction, and address, play different roles in the Silubium transaction system. Meanwhile, the four objects are correlated with one another and form a hierarchical relational structure. The design of SilkViser should enable users to immediately realize the independence and relevance of the four types of objects.

\textbf{DP2:} Integrating tabular and visual presentations. Most EUsers are accustomed to SilkViewer's textual and tabular interface. We should respect their habits and still show the majority of essential information fields in tabular forms. On this basis, we should adopt visualizations to present the newly extracted advanced information and illustrate important concepts and processes. This approach enables EUsers to smoothly transit from SilkViewer to SilkViser. 

\textbf{DP3:} Using familiar visual metaphors and lightweight interactions. SilkViser will be used by various users, and most of them lack experience in advanced visualization systems. SilkViser should consider their mental maps and use familiar metaphors and easy-of-use interactions to reduce difficulty in using the system. 

\begin{figure*}[htb]
	\centering
	\vspace{-0.2cm}  
	\setlength{\abovecaptionskip}{0.03cm}   
	\setlength{\belowcaptionskip}{-0.4cm}   
	\includegraphics[width=\textwidth]{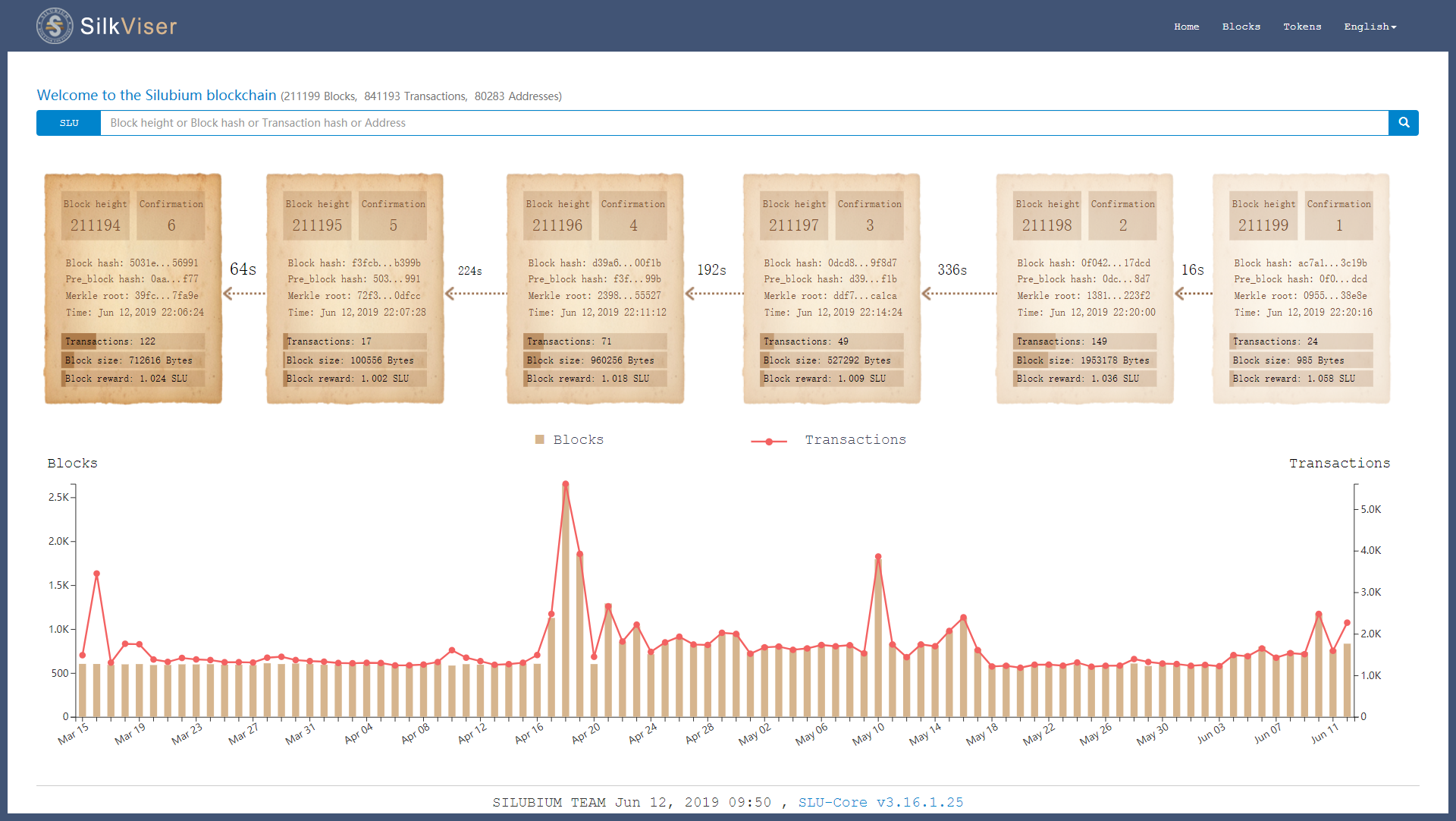}
	\caption{Visualization design of the blockchain page.}
	\label{figure2}
\end{figure*}
\section{SilkViser Design}\label{silkviser design}
In this section, we first introduce SilkViser's data processing and overall design, and then detail the visualization designs of each web page. \autoref{figure1} provides an overview of SilkViser.
\subsection{Data Processing}
We conduct a three-step data process for data abstraction. First, we identify the four types of objects. Second, we refine the essential information from basic data fields in accordance with the feedback gathered during the interviews. Lastly, we extract advanced information related to each object from the transaction data. The descriptions of the extracted essential and advanced information of the four types of objects are provided in the supplementary materials.
\subsection{Overall Interface Design}\label{overall interface design}
In the overall interface design, we answer several important design questions. The questions include the following: What is the preferred color scheme for the user interface? What visual experience do we hope the color scheme will convey to users? How many levels of functional web pages should be provided in SilkViser? What types of exploring mode should be used in the web pages?

We decide to use navy blue, sky blue, and aged paper yellow as the interface's main colors to convey a combinational sense of reliability, high-technology, and wealth in visual experience. Navy blue, which signifies the sea and evokes an emotion of extensity and reliability, is used as the frame color of the user interface. Sky blue, which symbolizes advanced technologies, is used to highlight functional buttons and navigational texts on the interface. Aged paper yellow is derived from gold and long-stored paper ledgers. It evokes a sense of wealth and history. 

In accordance with DP1, we decide to design four web pages: blockchain, block, transaction, and address pages. Each web page presents the information of a principal object and a number of correlated subordinate objects to form a hierarchical exploring mode. For example, the principal object of the block page is a certain block, and the correlated subordinate objects are the transactions packaged in the block. The principal object of the transaction page is a certain transaction, and the correlated subordinate objects are the addresses that participate in this transaction. In this manner, users can select an interesting subordinate object on a web page and then jump to the next-level web page that considers the interesting subordinate object as its principal object. Moreover, we provide a query exploring mode to enable users to directly access the web page that corresponds to a block, a transaction, or an address via keyword search. The two exploring modes can be used jointly during data viewing.
\subsection{Blockchain Page Design}\label{blockchain page design}
The blockchain page, as the homepage of SilkViser, presents the essential information of the blockchain (the principal object of this page) and the latest blocks (the subordinate objects of this page). This page is required to explain the concepts and processes related to blockchain and block objects (DR2) and to visualize the advanced information of the blockchain (DR3). 

\textbf{Layout design.} The blockchain page has three functional areas, as shown in \autoref{figure2}. A query operation area at the top of this page contains a query box entitled with a textual welcome message. The message displays the essential information of the blockchain, including the total numbers of blocks, transactions, and addresses on the blockchain. This query operation area is maintained in the same position as the three other web pages for convenient access to the query exploring mode. The middle area of this page is used for concept explanations and process illustrations. The bottom area is used for visualizing advanced information of Silkchain.

\begin{figure}[htb]
	\centering
	\setlength{\abovecaptionskip}{-0.05cm}   
	\setlength{\belowcaptionskip}{-0.3cm}   
	\includegraphics[width=0.9\columnwidth]{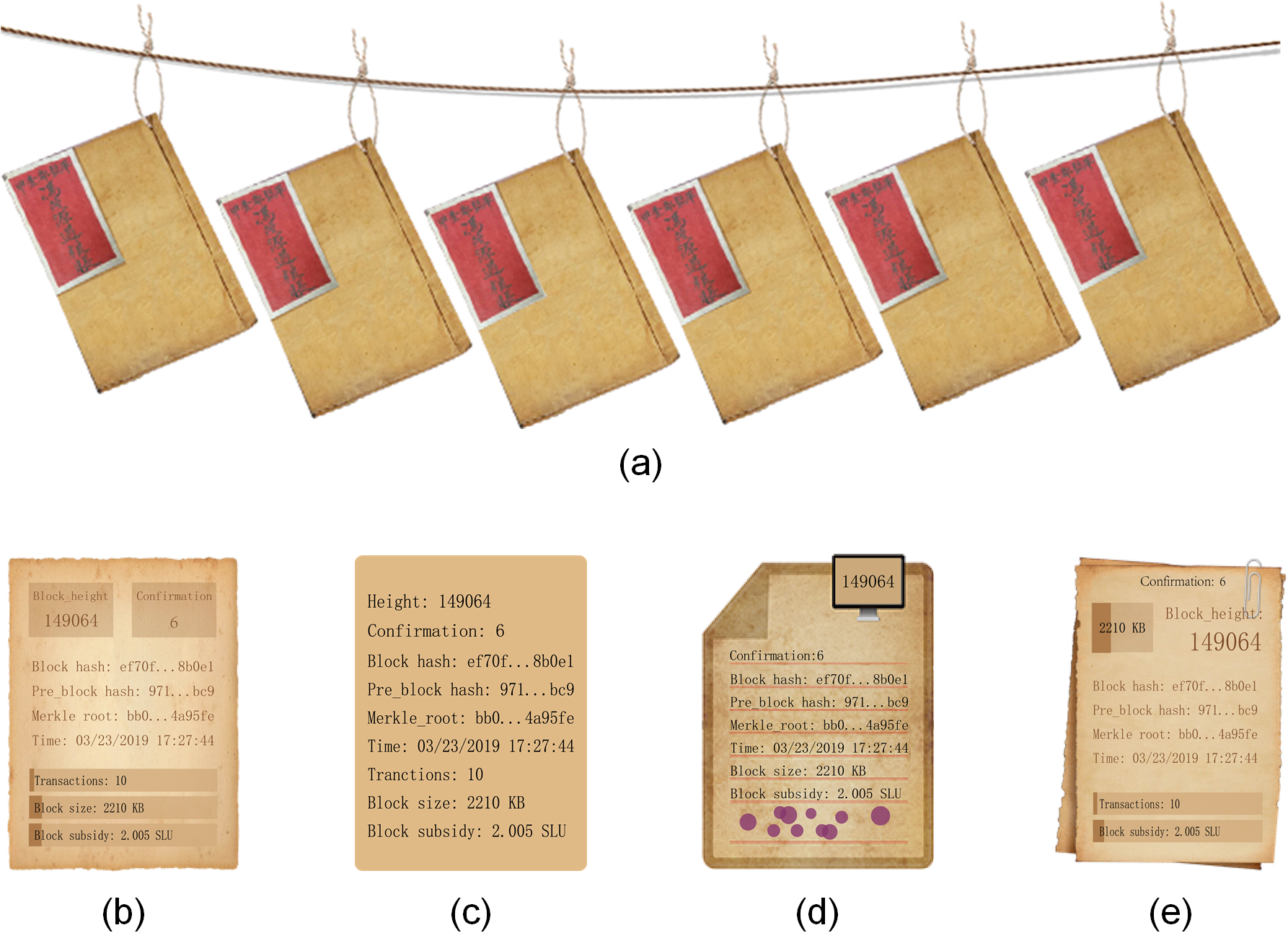}
	\caption{Visualization designs of the blockchain and blocks: (a) visual metaphor of linked paper ledgers, (b) final design of block glyph, and (c$  -  $e) alternative designs of block glyph.}
	\label{figure3}
\end{figure}

\textbf{Visual metaphor design for the blockchain.} Visually explaining a set of concepts (e.g., the blockchain, blocks, and their relationship) and processes (e.g., chain-forming and block confirming) related to the blockchain and block objects in the middle area of the blockchain page is challenging. Inspired by the traditional manner of accounting, we adopt the metaphor of a string of paper ledgers (\autoref{figure3}(a)) to address this design challenge. Before computers were used for accounting, a company or an organization had to write down the amounts of money it spent and received on paper ledgers. These ledgers were then strung together in chronological order and hung in the ledger room for easy access. The use of this metaphor design has two major benefits. First, using this metaphor can evoke natural associations between the users' mental map and domain knowledge of cryptocurrencies (DP3). For example, a string of paper ledgers naturally conveys a scene of the chain, and a paper ledger on a string can be easily associated with a block with limited memory space for recording transactions. Second, aged paper ledgers provide a sense of history and reliability.

As shown in \autoref{figure2}, we use a yellow square, which is similar to an aged paper ledger, to represent a block. A dotted line with an arrow between two blocks, which is similar to the string used in paper ledgers, represents the successive relationship between blocks. The arrow pointing to the left indicates that a block always has a connected previous block. The length of a line denotes the time interval for generating two blocks. When a new block is generated, the leftmost square in the page is moved out of the screen and a new square is linked to the rightmost side. These design details can intuitively explain that the blockchain is formed by linking continuously generated blocks, which are represented as public ledgers for recording transaction information. We set that up to six blocks can be shown in the page. Displaying the latest blocks (usually 5 \~{} 20 blocks) in textual forms at homepages is a common way of blockchain explorers. This way can satisfy the requirement of most clients. We use the setting of displaying the latest 6 blocks because it benefits the illustration of the six-confirmation concept and process and also avoids visual clutter caused by showing too many blocks.

\textbf{Visual glyph design for a block:} In the visual metaphor, a square glyph is designed to represent the essential information and concepts related to the block object. As shown in \autoref{figure3}(b), the block glyph has three parts from top to bottom. The first part includes two conspicuous boxes at the top. The text on the left box shows the height of a block on Silkchain (i.e., the sequence number in the entire chain), which is similar to an ID label on the cover of a paper ledger. The text on the right box indicates the current confirmation number of a block. This confirmation number benefits the illustration of the six-confirmation process. The second part shows a list of texts in the middle that indicate the generation time of a block, the block hash, and the previous block hash. The third part includes three bands with embedded texts and filled areas that represent three important statistical information of a block: number of transactions, block size, and block reward. Moreover, the color intensity of a block is a double encoding of the six-confirmation process. Dark color indicates a high amount of confirmation.

\autoref{figure3}(c$  -  $e) present several design alternatives that have been evaluated. The plain textual and tabular form of the first alternative (\autoref{figure3}(c)) is monotonous and unfocused. The second alternative (\autoref{figure3}(d)) introduces several visual elements, including a dog-ear at the top left corner to vivify the interface, a computer icon at the top right corner to show the height of a block, and a scatterplot at the bottom to represent transactions. However, the dog-ear and computer icon not only waste space but also do not concur with the metaphor's retro style. The scatterplot can be subject to serious visual clutter if a block has numerous transactions. The third alternative (\autoref{figure3}(e)) uses burring edges to evoke a sense of retro style, which is maintained in the final design. However, the paper clip at the top right corner may lead to a misunderstanding that the glyph presents multiple blocks. The final design (\autoref{figure3}(b)) has three advantages. First, the overall design is simple and intuitive. Second, the three-part layout is visually coordinated with the layout of the home page. Third, the three parts use diverse visual elements, making it easy for users to recognize different information. Other visualization design alternatives in the blockchain page are provided in the supplementary materials.

\textbf{Visualization of advanced information of the blockchain.} The bottom area of the blockchain page visualizes the advanced information of Silkchain. EUsers want to know the daily generation trends of blocks and transactions in the last 3 months. We use a compounded line and bar chart with double \textit{y}-axes to satisfy the requirement. As shown in \autoref{figure2}, the \textit{x}-axis presents the last 3 months by day, and the \textit{y}-axes on the left and right represent the number of generated blocks and transactions per day, respectively. The line inside the chart represents the transaction generation trend, and the bars inside the chart denote the block generation trend. Line and bar charts are commonly used to show trends and are familiar to users (DP3). Presenting line-style and bar-style trends in one chart is helpful for users to distinguish between the two trends and compare them in the same period. As shown in \autoref{figure2}, in the 3 months from March 15 to June 12, the numbers of blocks and transactions generated daily on Silkchain were generally stable, and their generation trends were highly consistent. The two distinct peaks on April 18 and May 10 were due to the sudden breakout of the cryptocurrency exchange market when many dormant traders became active.

\subsection{Block Page Design}\label{block page design}
The block page aims to provide users with three types of information related to a block, namely, the essential information of the block (i.e., the principal object of this page), the list of transactions (i.e., the subordinate objects of this page) in the block, and the explanation of the one-to-many relationship between block and transaction objects (DR2) and the visualization of the advanced information of a block (DR3), including the distributions of fees, the number of addresses, and the sizes of transactions in a block.

\begin{figure}[htb]
	\centering
\vspace{-0.2cm}  
	\setlength{\abovecaptionskip}{-0.01cm}   
	\setlength{\belowcaptionskip}{-0.2cm}   
	\includegraphics[width=\columnwidth]{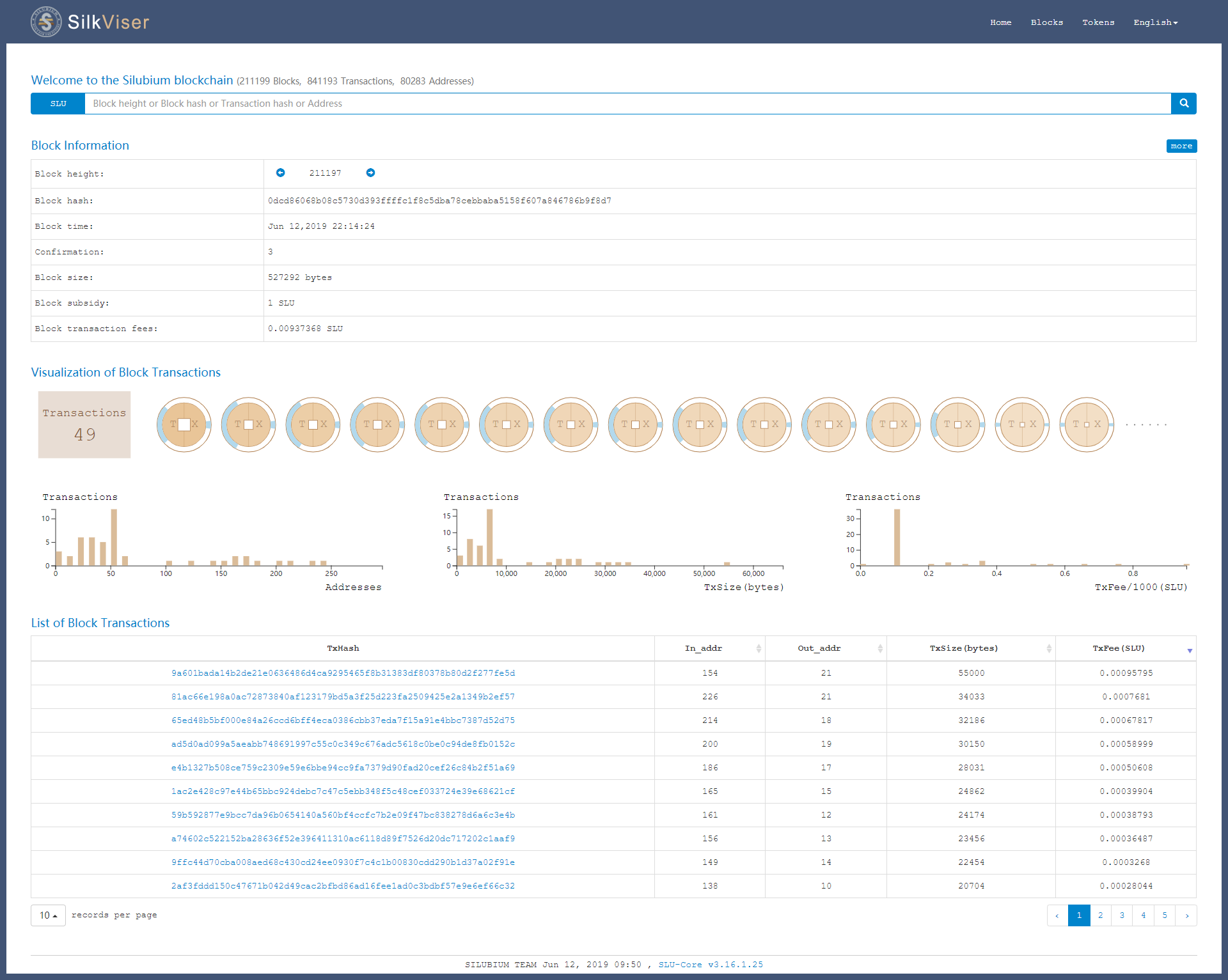}
	\caption{Visualization design of the block page.}
	\label{figure4}
\end{figure}

\textbf{Layout design.} The block web page is vertically divided into three functional areas to present the three types of information, as shown in \autoref{figure4}. The top area uses a vertical two-column table to display the names and values of the essential information fields of a block. The bottom area lists all the transactions in the block in a paging horizontal table with five columns, namely, the ID of a transaction (TxHash), the number of inputs (In\_addr) and outputs (Out\_addr) in a transaction, the memory size occupied by a transaction (TxSize), and the fee of a transaction (TxFee). The middle area visually presents the third type of information.

This three-area layout design has two advantages. First, this design integrates tabular and visual forms in information presentation, and thus, is in line with DP2 and also helpful in immediately distinguishing the three types of information. Second, it is conducive to the expression of independence and relevance of block and transaction objects (DP1). The visualization area at the middle is a natural divider that allows users to intuitively feel that the two tabulations at the top and bottom areas represent two different types of objects. Meanwhile, the middle area uses visualizations to explain the one-to-many relationship between the block and transaction objects.

\textbf{Visualization and interaction design of the block page.} The middle area is divided into the upper and lower sub-areas. The upper sub-area uses a visual design composed of a square and a string of circles. The square at the far left represents a block and displays the number of transactions in the block, indicating the one-to-many relationship between the block and transaction objects. The string of circles is at the right side of the square, and each circle represents a transaction, explaining that many transactions occur in the block. We provide an option to set the maximum number of circles displayed, and the appropriate setting for most common desktop displays is 15 circles. That is, if more than 15 transactions transpired in this block, then the string of circles is followed by an ellipsis at the end.

The lower sub-area presents the distributions of fees, the number of addresses, and the sizes of transactions in a block. Such advanced information is of high concern among EUsers. For example, a Silubium administrator may check if a small number of transactions in a block are taking up most of the block's storage space. A miner wants to know whether the mining reward comes primarily from a few transactions. We use three classic bar charts to visualize the three distributions (DP3). Their \textit{y}-axes represent the number of transactions, and their \textit{x}-axes are the number of addresses of a transaction, transaction size, and transaction fee.

EUsers have the requirements of finding and comparing transactions of interest in a block. For example, miners are concerned with identifying which transactions provide higher transaction fees when packaging a block, and system administrators are interested in determining which transactions consume the majority of the memory space of a block or involve a large number of addresses. In this page, we provide users with two simple interactions, namely, sorting and brushing (DP3), and then design a transaction glyph (see the next paragraph) to satisfy this requirement. Users can click the mouse on any column header of the transaction table at the bottom area to sort transactions in ascending, descending, or random order based on the values in the clicked column. Users can also brush a section on the \textit{x}-axis of any bar chart in the middle area, and the transactions that satisfy the filter conditions are displayed in the transaction table. The two interactions also affect the transactions represented by the circles in the middle area. As shown in \autoref{figure4}, the transactions in a block are listed in descending order in accordance with TxFee. The top 15 transactions are shown at the upper sub-area, with descending order from left to right.  

\begin{figure}[htb]
	\centering
	\setlength{\abovecaptionskip}{0.1cm}   
	\setlength{\belowcaptionskip}{-0.2cm}   
	\includegraphics[width=0.9\columnwidth]{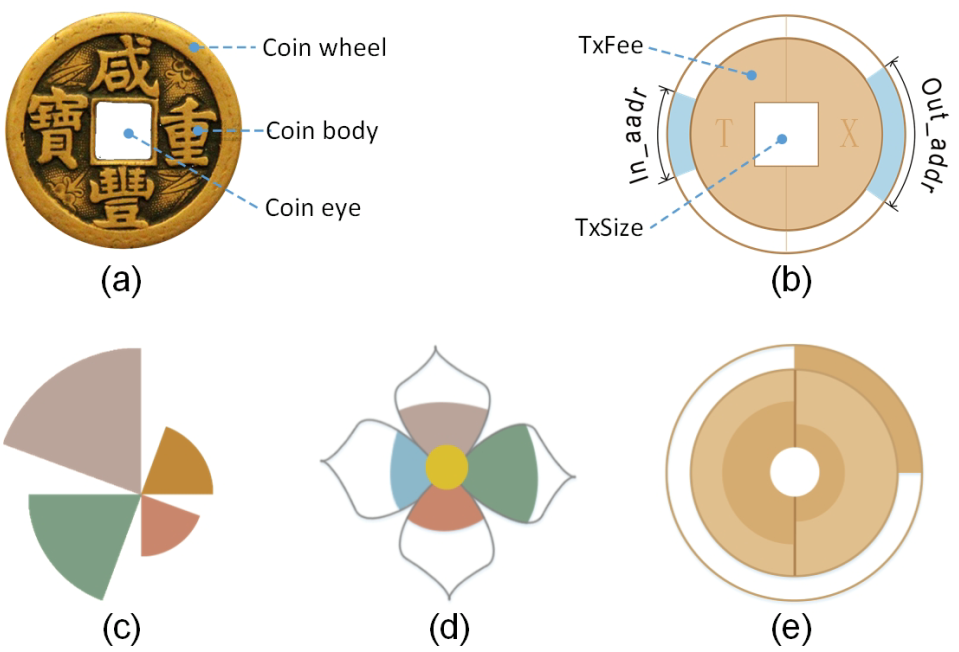}
	\caption{Glyph design of a transaction: (a) copper coin, (b) coin-like glyph, and (c$ - $e) three design alternatives, where TxFee and TxSize refer to the fee of a transaction and the memory size occupied by a transaction respectively; In\_addr and Out\_addr mean the number of inputs and outputs in a transaction respectively.}
	\label{figure5}
\end{figure}

\textbf{Glyph design of the block page.} A transaction glyph is designed in the middle area of this page to form a circle that represents transaction information for transaction comparison. The two key points in the glyph design are the simultaneous presentation of four essential information fields of a transaction (i.e., In\_addr, Out\_addr, TxSize, and TxFee) and the coordinated visual sense with the overall interface design. After producing several iterative designs and in-depth discussions, a visual metaphor of copper coins from ancient China is adopted for this glyph design. \autoref{figure5}(a) shows a copper coin minted and circulated during the reign of Emperor Xianfeng (1831--1861). It has three components: a coin wheel, a coin body with mint marks, and a coin eye used for tying up coins with a string for easy carrying. \autoref{figure5}(b) illustrates the glyph design. The four information fields are encoded by utilizing the three components of a copper coin with three types of visual elements. We divide the coin wheel into two fillable rings along the vertical centerline. The left and right rings can be filled to a certain percentage with blue, representing In\_addr and Out\_addr, respectively. A large filled percentage indicates a high number of input or output addresses. We use the color of the coin body to encode Txfee. Dark bronze-yellow color indicates a high transaction fee. We associate the size of the coin eye with TxSize. A large coin eye indicates a large TxSize. In addition, the letters T and X fixed on the coin body indicate that this glyph represents a transaction. As shown in \autoref{figure4}, comparing the first and second coin glyphs, the first has a larger coin eye, but the second coin has a larger left ring, indicating that the first transaction occupies more memory in the block and the second transaction has more input addresses. 

This glyph design has four advantages. First, the financial property of the coins naturally indicates the commercial purpose of cryptocurrency transactions. Then, the bronze-yellow copper coin is compatible with aged paper yellow. Third, the circular shape of the coin satisfies the overall design idea of SilkViser, wherein a square represents a block and a circle represents a transaction. The coin exhibits good visual stability in facilitating transaction comparison. Lastly, a copper coin has diverse structural components that can be easily utilized to encode multiple types of information. \autoref{figure5}(c$  -  $e) present several design alternatives. The first alternative uses a nightingale rose diagram with four sectors to encode four information fields. However, the overall size of a rose is unstable, distracting users' attention when displaying roses with different sizes. The second alternative uses a flower with four petals to represent a transaction. The four petals have the same fixed size, and the filled percentage of a petal indicates the value of the corresponding information field. This alternative is helpful for information recognition and transaction comparison, but inner or imagery connection is lacking between flowers and cryptocurrency transactions. We also propose a straightforward circular design, as shown in \autoref{figure5}(e). It is close to the final design. More design alternatives are provided in the supplementary materials.
\subsection{Transaction Page Design}\label{transaction page design}
\begin{figure}[htb]
	\centering
	\setlength{\abovecaptionskip}{-0.01cm}   
	\setlength{\belowcaptionskip}{-0.3cm}   
	\includegraphics[width=\columnwidth]{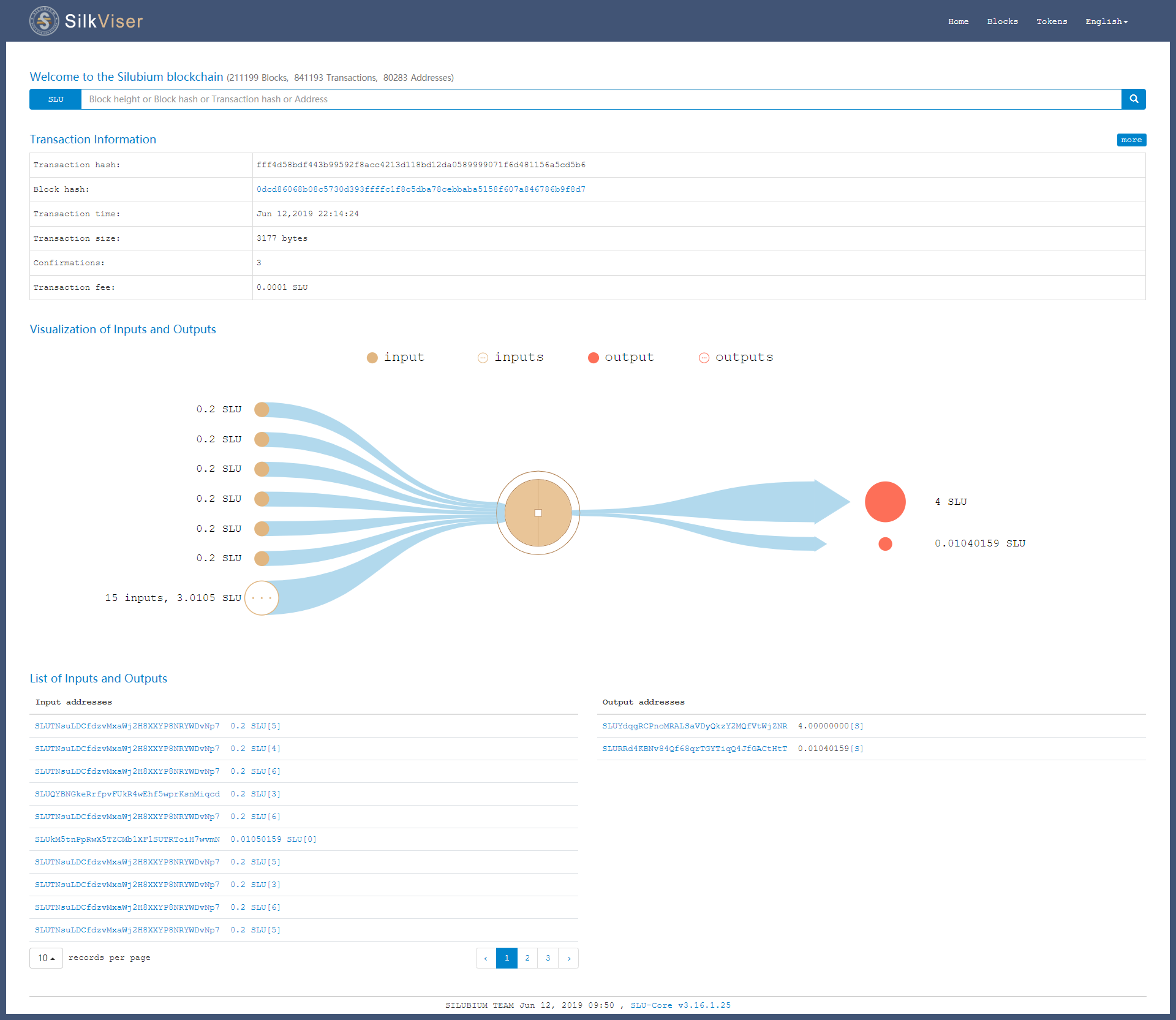}
	\caption{Visualization design of the transaction page.}
	\label{figure6}
\end{figure}

The transaction page presents three types of information. The first type is the essential information of a transaction (i.e., the principal object of this page). The second type is the input and output addresses that participate in the transaction (i.e., the subordinate objects of this page). The third type explains relevant concepts (DR2) and visualizes the advanced information of a transaction (DR3). This page adopts a three-area layout similar to the block page (DP1 and DP2), as shown in \autoref{figure6}. The top area uses a vertical table to display the essential information of a transaction. The bottom area uses two juxtaposed paging horizontal tables to show the participating input and output addresses. Users can click the mouse on any column header of the tables to sort the addresses and on any address to proceed to its corresponding page (DP3).

The middle area uses visualizations to (1) explain that a transaction may involves multiple input and output addresses, (2) illustrate the concept of coin mixing, and (3) visualize the primary input and output addresses that receive and send the majority of SLU in the transaction. We propose a coin-Sankey diagram \cite{F6,F7}, which integrates the coin glyph design in the block page into traditional Sankey diagrams, to accomplish the three visual design tasks. As shown in \autoref{figure6}, the center of the coin-Sankey diagram is a copper coin glyph that represents the transaction on this page. Its design details are consistent with that of the coin glyph in the block page. The circles on the left and right sides of the coin glyph represent the inputs and outputs participating in the transaction, respectively. Using the Sankey diagrams as a reference, a circle is connected to the wheel of the coin glyph by a ribbon. A large circle radius and a wide ribbon indicate that the relevant address sends or receives a high amount of SLU in the transaction. The SLU amount of each address appears in the textual form near the corresponding circle. Only the top six inputs and outputs in terms of SLU amount can be displayed to avoid visual clutter caused by simultaneously showing numerous addresses. The remaining inputs and outputs are merged into the left-last and right-last circles, respectively, which are specially marked by three small dots inside. As shown in \autoref{figure6}, 21 inputs exist in the transaction, and 6 inputs are shown in the diagram. All the inputs have the same SLU amount, which is 0.2 SLU. The 15 other addresses, which are compressed and displayed in the last circle with dots, provide a total of 3.0105 SLU. This transaction has two outputs, one of which receives the majority amount of SLU.

This coin-Sankey diagram design has many advantages. The Sankey diagram is a common visual diagram (DP3). The use of the coin glyph provides continuity of visual perception from the block page to the transaction page. A coin glyph simultaneously displayed with multiple circles naturally explains the one-to-many relationship between the transaction and address objects. The limited display of circles allows the primary inputs and outputs to be easily recognized and selected to proceed to the corresponding address pages. Sankey-style ribbons start from the left circles (representing input addresses), converge to the left wheel of the coin glyph, disperse from the right wheel, and eventually end at the right circles (representing output addresses). This design illustrates the concept of coin mixing and intuitively depicts the flow direction of SLU in a transaction.

\subsection{Address Page Design}\label{address page design}
The address page is the last level of SilkViser pages. It presents the essential information of an address; the transactions in which this address participated; and the advanced statistical information of the address, including the balance variant trend and transaction participation trend of the address in the last 30 days. This page also adopts a three-area layout similar to the block page. The top area uses a vertical table to display the essential information of the address. The bottom area uses a paging horizontal table to list the transactions in which the address has participated. Users can select any transaction to proceed to the corresponding transaction page. The middle area is the visualization area that depicts the two trends. As shown in \autoref{figure1}(d), a line-bar compounded chart with double \textit{y}-axes is designed to visualize the two trends, which are similar to the design of trend visualization in the blockchain page.
\section{User Study}\label{user study}
We conducted a user study to evaluate the usability and effectiveness of SilkViser. We invited a number of EUsers and NUsers to complete an objective questionnaire and a subjective questionnaire by using SilkViser to browse the Silubium transaction data. Examples of the subjective and objective questionnaires and usage cases of NUsers and EUsers are provided in the supplementary materials.
\begin{table*}[htb]
\caption{Results of mean accuracy, mean time, mean easiness score, and mean confidence score in solving each objective question. The 25 questions are categorized from three perspectives, namely, web page, information type (ES: essential information, AD: advanced information, and CP: concept and process), and exploring mode (H: hierarchical exploring mode and Q: query exploring mode). The colors indicate that significant differences exist between the EUsers and the NUsers. The winners are shown in dark blue, and the losers are shown in light blue. Using Q2 as an example, significant differences are found between the EUsers and the NUsers in terms of time, easiness, and confidence.}
\label{tab:table1}
\scriptsize
\renewcommand{\arraystretch}{1.25}
\newcommand{\tabincell}[2]{\begin{tabular}{@{}#1@{}}#2\end{tabular}}
\addvbuffer[0pt -11pt]{
\begin{tabular}{|m{1.1cm}<{\centering}|m{1.4cm}<{\centering}|m{1.4cm}<{\centering}|m{1.3cm}<{\centering}|m{0.9cm}<{\centering}|m{0.9cm}<{\centering}|m{0.9cm}<{\centering}|m{0.9cm}<{\centering}|m{0.9cm}<{\centering}|m{0.9cm}<{\centering}|m{0.9cm}<{\centering}|m{0.9cm}<{\centering}|}
\hline
\multirow{2}*{\textbf{Question}}&\multirow{2}*{\textbf{Web page}}&\multirow{2}*{\textbf{\tabincell{c}{Information \\ Type}}}&\multirow{2}*{\textbf{\tabincell{c}{Exploring \\ Mode}}}&\multicolumn{2}{c|}{\textbf{Accuracy}}&\multicolumn{2}{c|}{\textbf{Time(s)}}&\multicolumn{2}{c|}{\textbf{Easiness}}&\multicolumn{2}{c|}{\textbf{Confidence}}\\ 
\cline{5-12}
&&&&EUsers&NUsers&EUsers&NUsers&EUsers&NUsers&EUsers&NUsers\\
\hline
Q1&\multirow{7}*{\tabincell{c}{Blockchain \\ page}}&ES&&0.95&0.87&22.55&28.52&4.14&4.09&4.32&4.22
\\
\hhline{-~|-|-|-|-|-|-|-|-|-|-|}
Q2&&CP&&0.95&0.78& 
\cellcolor[rgb]{.208,.488,.831}\color{white}5.36 & 
\cellcolor[rgb]{.411,.643,.851}\color{white}12.96&
\cellcolor[rgb]{.208,.488,.831}\color{white}4.36&
\cellcolor[rgb]{.411,.643,.851}\color{white}3.57&
\cellcolor[rgb]{.208,.488,.831}\color{white}4.23&
\cellcolor[rgb]{.411,.643,.851}\color{white}3.52
\\ 
\hhline{-~|-|-|-|-|-|-|-|-|-|-|}
Q3&&CP&&
0.84&0.87&
\cellcolor[rgb]{.208,.488,.831}\color{white}14.23&
\cellcolor[rgb]{.411,.643,.851}\color{white}17.39&3.73&4.04&3.59&4.17\\
\cline{1-1}\cline{3-12}
Q4&&AD&&0.91&0.96&10.00&16.74&4.09&3.87&4.00&3.96\\ 
\hhline{-~|-|-|-|-|-|-|-|-|-|-|}
Q5&&AD&&0.95&0.96&
\cellcolor[rgb]{.208,.488,.831}\color{white}11.05&
\cellcolor[rgb]{.411,.643,.851}\color{white}21.91&4.14&3.96&4.05&4.00\\ 
\hhline{-~|-|-|-|-|-|-|-|-|-|-|}
Q6&&ES, CP&&0.91&0.88&
\cellcolor[rgb]{.208,.488,.831}\color{white}50.95&
\cellcolor[rgb]{.411,.643,.851}\color{white}82.04&3.91&3.65&4.09&3.74\\
\hhline{-~|-|-|-|-|-|-|-|-|-|-|}
Q7&&ES&&0.89&0.82&
\cellcolor[rgb]{.208,.488,.831}\color{white}32.86&
\cellcolor[rgb]{.411,.643,.851}\color{white}51.00&
\cellcolor[rgb]{.208,.488,.831}\color{white}4.14&
\cellcolor[rgb]{.411,.643,.851}\color{white}3.48&4.14&3.52\\
\hline
Q8&\multirow{10}*{Block page}&ES, CP&H&1.00&0.96&
\cellcolor[rgb]{.208,.488,.831}\color{white}48.00&
\cellcolor[rgb]{.411,.643,.851}\color{white}85.17&4.27&3.78&4.27&3.83\\ 
\hhline{-~|-|-|-|-|-|-|-|-|-|-|}
Q9&&AD&H&0.64&0.55&64.27&74.43&
\cellcolor[rgb]{.208,.488,.831}\color{white}4.23&
\cellcolor[rgb]{.411,.643,.851}\color{white}3.03&
\cellcolor[rgb]{.208,.488,.831}\color{white}4.64&
\cellcolor[rgb]{.411,.643,.851}\color{white}3.30\\
\hhline{-~|-|-|-|-|-|-|-|-|-|-|}
Q10&&ES&H&0.91&0.74&
\cellcolor[rgb]{.208,.488,.831}\color{white}41.36&
\cellcolor[rgb]{.411,.643,.851}\color{white}80.87&
\cellcolor[rgb]{.208,.488,.831}\color{white}4.18&
\cellcolor[rgb]{.411,.643,.851}\color{white}3.35&
\cellcolor[rgb]{.208,.488,.831}\color{white}4.32&
\cellcolor[rgb]{.411,.643,.851}\color{white}3.26
\\ 
\hhline{-~|-|-|-|-|-|-|-|-|-|-|}
Q11&&ES&H&0.82&0.80&
\cellcolor[rgb]{.208,.488,.831}\color{white}45.91&
\cellcolor[rgb]{.411,.643,.851}\color{white}83.39&4.02&3.86&3.98&3.65\\ 
\hhline{-~|-|-|-|-|-|-|-|-|-|-|}
Q12&&ES&H&0.86&0.78&
\cellcolor[rgb]{.208,.488,.831}\color{white}25.73&
\cellcolor[rgb]{.411,.643,.851}\color{white}46.13&4.05&3.65&4.00&3.57\\ 
\hhline{-~|-|-|-|-|-|-|-|-|-|-|}
Q17&&ES, CP&Q&1.00&0.99&
\cellcolor[rgb]{.208,.488,.831}\color{white}45.50&
\cellcolor[rgb]{.411,.643,.851}\color{white}60.39&4.36&4.00&4.45&4.26\\
\hhline{-~|-|-|-|-|-|-|-|-|-|-|}
Q18&&AD&Q&0.68&0.64&29.41&27.96&3.82&3.52&3.73&3.43\\ 
\hhline{-~|-|-|-|-|-|-|-|-|-|-|}
Q19&&ES&Q&0.95&0.74&
\cellcolor[rgb]{.208,.488,.831}\color{white}29.68&
\cellcolor[rgb]{.411,.643,.851}\color{white}55.27&
\cellcolor[rgb]{.208,.488,.831}\color{white}4.27&
\cellcolor[rgb]{.411,.643,.851}\color{white}3.48&
\cellcolor[rgb]{.208,.488,.831}\color{white}4.36&
\cellcolor[rgb]{.411,.643,.851}\color{white}3.48\\ 
\hhline{-~|-|-|-|-|-|-|-|-|-|-|}
Q20&&AD&Q&0.68&0.61&
\cellcolor[rgb]{.208,.488,.831}\color{white}33.68&
\cellcolor[rgb]{.411,.643,.851}\color{white}55.57&4.12&3.77&
\cellcolor[rgb]{.208,.488,.831}\color{white}3.87&
\cellcolor[rgb]{.411,.643,.851}\color{white}3.21\\ 
\hhline{-~|-|-|-|-|-|-|-|-|-|-|}
Q21&&AD&Q&0.86&0.78&29.82&28.17&4.27&3.78&
\cellcolor[rgb]{.208,.488,.831}\color{white}4.32&
\cellcolor[rgb]{.411,.643,.851}\color{white}3.74\\ 
\hline
Q13&\multirow{4}*{\tabincell{c}{Transaction \\  page}}&ES&H&0.89&0.94&
\cellcolor[rgb]{.208,.488,.831}\color{white}74.27&
\cellcolor[rgb]{.411,.643,.851}\color{white}90.30&4.09&3.48&
\cellcolor[rgb]{.208,.488,.831}\color{white}4.27&
\cellcolor[rgb]{.411,.643,.851}\color{white}3.61\\ 
\hhline{-~|-|-|-|-|-|-|-|-|-|-|}
Q14&&AD&H&0.89&0.78&70.00&71.33&3.68&3.10&3.77&2.95\\
\hhline{-~|-|-|-|-|-|-|-|-|-|-|}
Q22&&ES&Q&0.87&0.83&90.23&83.35&
\cellcolor[rgb]{.208,.488,.831}\color{white}4.18&
\cellcolor[rgb]{.411,.643,.851}\color{white}3.30&
\cellcolor[rgb]{.208,.488,.831}\color{white}4.18&
\cellcolor[rgb]{.411,.643,.851}\color{white}3.39\\ 
\hhline{-~|-|-|-|-|-|-|-|-|-|-|}
Q23&&AD&Q&0.93&0.96&38.05&42.26&4.18&3.65&4.23&3.61\\
\hline
Q15&\multirow{4}*{Address page}&ES&H&0.84&0.83&87.55&93.65&3.95&3.65&4.05&3.70\\ 
\hhline{-~|-|-|-|-|-|-|-|-|-|-|}
Q16&&AD&H&0.57&0.33&
\cellcolor[rgb]{.208,.488,.831}\color{white}41.91&
\cellcolor[rgb]{.411,.643,.851}\color{white}45.00&3.64&3.22&3.73&3.17\\ 
\hhline{-~|-|-|-|-|-|-|-|-|-|-|}
Q24&&ES&Q&0.97&0.87&
\cellcolor[rgb]{.208,.488,.831}\color{white}62.77&
\cellcolor[rgb]{.411,.643,.851}\color{white}76.83&4.14&3.52&4.23&3.65\\ 
\hhline{-~|-|-|-|-|-|-|-|-|-|-|}
Q25&&AD&Q&0.61&0.59&46.05&49.43&3.77&3.43&3.86&3.30\\ 
\hline 
\end{tabular}}
\end{table*}
\subsection{Participants, Data, and Apparatus}\label{participants, data, and apparatus}
We recruited 23 NUsers (14 females and 9 males; aged 19--25 years; median: 20 years) and 22 EUsers (6 females and 16 males; aged 23--34 years; median: 28 years) for the formal study. The NUsers are undergraduate students majoring in finance and economics. They have no actual cryptocurrency transaction experience, but are highly interested in cryptocurrencies. The EUsers are SLU clients with at least 3 months of Silubium transaction experience and more than 10 transaction records. They exhibit a strong motivation to try out new tools. Each participant was compensated with 400 SLU.
 
We provided two experimental datasets, which were derived from the real-world Silubium transaction data in different periods. One period was from June 12, 2019 to September 10, 2019, with a total of 856,192 transactions as the training dataset. The other period was from March 15, 2019 to June 12, 2019, with a total of 841,193 transactions as the formal experiment dataset. The experimental apparatus was a Dell OptiPlex 7050 and a 23.8-inch screen with a resolution of 1920 $\times$ 1080. SilkViser was preloaded on the Google Chrome browser. The participants used a standard wired mouse and wired keyboard for interactions.

\subsection{Experiment Procedure}\label{experiment procedure}
The formal study had four sessions. In this section, we use a certain participant as an example to introduce the procedure.

\textit{\textbf{Introduction and demonstration.}} For a certain participant, the instructor explained the purpose and procedure of the experiment and provided background knowledge. Then, the instructor introduced SilkViser's interface and operations.

\textit{\textbf{Review and trial.}} The instructor guided the participant to use SilkViser to browse the training dataset and answer the questionnaires. The participant was allowed to ask questions.

\textit{\textbf{Formal experiment and questionnaires.}} The participant was asked to complete the objective questionnaire using the formal experiment dataset. The objective questionnaire contains 25 questions. The 25 questions can be categorized from three perspectives, namely, web page, information type, and exploring mode, as shown in \autoref{tab:table1}. For the web page, the questions were divided into four groups under the four-level web pages. For information type, the questions were divided into three groups: essential information, advanced information, and related concepts and processes. For exploring mode, the questions were divided into two groups, namely, the hierarchical and query exploring modes. After completing each objective question, the participant was required to use a five-point Likert scale ranging from 5 (strongly difficult/strongly confident) to 1 (strongly easy/strongly diffident) to measure the difficulty of this question and their confidence in answering. After completing the objective questionnaire, the participant was asked to fill the subjective questionnaire with 16 subjective questions (SQs). The first six SQs was used to rate the overall feelings toward SilkViser, including satisfaction, intuitiveness, informativeness, ease of use, effectiveness, and willingness. SQ7--SQ9 were used to rate the designs of the exploring modes, layout, and interactions. The other SQs were about the visualization designs of each web page.

\textit{\textbf{Interview.}} After the formal study, we interviewed with the participant. He/she was encouraged to state the problems he/she encountered and his/her feelings as much as possible.
\subsection{Analysis of Results}\label{analysis of results}
\subsubsection{Analysis of objective questionnaire results}
We completely recorded two objectives and two subjective metrics as the experimental results of the objective questionnaire. The objective metrics are the accuracy and time of each participant in completing each question. The subjective metrics are the easiness and confidence scores rated by each participant after completing each question. Our analytical approach included three aspects. (1) We calculated the means and standard deviations of the metrics of all the participants with respect to the questions grouped by the web page. The result is presented in \autoref{tab:table2}. (2) We calculated the means and standard deviations of all the metrics of all the participants with respect to the questions categorized by exploring mode. The result is provided in \autoref{tab:table3}. (3) We calculated the means and standard deviations of the metrics of two types of users (NUsers and EUsers) on each question. We also examined whether significant differences existed between the two types of users in the four metrics, as shown in \autoref{tab:table1}. The significance analysis was conducted using nonparametric Kruskal--Wallis because the objective questionnaire results did not follow a normal distribution in accordance with the Shapiro--Wilk testing results. All the tests were performed under the level \textit{p} = 0.05 for determining statistical significance.  

\textit{\textbf{(1) Result analysis by web page}}

Objective questions Q1--Q7 were mainly about the blockchain page. As shown in the first row of \autoref{tab:table2}, the participants obtained relatively high accuracy ($\mu$ = 0.90) with high confidence ($\mu$ = 3.97, $\sigma$ = 0.14) on the seven questions, reflecting that the design of the blockchain page can effectively help them understand concepts and processes related to the blockchain and block objects and recognize the essential and advanced information of Silkchain and newly generated blocks. Moreover, the participants in the blockchain page performed the best in terms of the four metrics among the four web pages, as indicated in \autoref{tab:table2}. Among the 45 participants, 36 mentioned that the blockchain page was highly impressive. Moreover, 28 participants said that they liked the paper ledger metaphor and the block glyph design, and 18 EUsers appreciated that the trend information can be directly viewed.
\begin{table}[!h]
	\scriptsize
	\renewcommand\tabcolsep{8.5pt}  
	\renewcommand\arraystretch{1.2} 
	\centering
	\vspace{0.1cm}  
	\caption{Objective questionnaire results in terms of mean accuracy, mean time, mean easiness score, and mean confidence score by web page.}
	\label{tab:table2}
\addvbuffer[0pt -8pt]{
	\begin{tabular}{m{2cm}<{\centering}m{1cm}<{\centering}m{0.5cm}<{\centering}m{0.8cm}<{\centering}m{1cm}<{\centering}}
\hline
Web page & Accuracy & Time &	Easiness &	\textls[-50]{Confidence}\\
\hline
Blockchain page &	0.90	& 27.10 &	3.93 &	3.97\\

Block page &	0.80 &	47.81&	3.85 &	3.83\\

Transaction page &	0.89&	70.04 &	3.74 &	3.80\\

Address page &	0.70 &	62.97 &	3.66&	3.71\\
\hline
\end{tabular}
}
\end{table}

Q8--Q12 and Q17--Q21 were mostly related to the block page. As shown in \autoref{tab:table2}, the participants obtained good accuracy ($\mu$ = 0.80) and rated a high confidence score ($\mu$ = 3.83, $\sigma$ = 0.22) on the 10 questions. This result reflected that the designs of the block page can help participants recognize block-related information and understand the one-to-many relationship between the block and transaction objects. Furthermore, the accuracies of Q9 ($\mu$ = 0.60), Q18 ($\mu$ = 0.66), and Q20 ($\mu$ = 0.65) were considerably lower than those of the other block page questions, as indicated in \autoref{tab:table1}. Our observations and the participants' feedback indicate that this result was mainly caused by imperfect question designs. Q9 and Q18 tested whether the participants can accurately identify three distributions related to all the transactions in the latest block. However, the latest block constantly changed and some participants encountered blocks with particularly ambiguous distributions. Q20 asked the participants to find a transaction that involved a large number of addresses in a given block. However, we failed to mention that addresses should simultaneously include inputs and outputs.

As shown in \autoref{tab:table2}, the participants obtained relatively good accuracy ($\mu$ = 0.89) with high confidence ($\mu$ = 3.80, $\sigma$ = 0.14) on the four questions (Q13, Q14, Q22, and Q23) related to the transaction page. The coin-Sankey diagram received highly positive comments from the participants. They agreed that this diagram can clearly explain the relationship between the transaction and address objects and help users identify important inputs and outputs. 

Among the four web pages, the address page gained the worst accuracy ($\mu$ = 0.70, $\sigma$ = 0.14), as indicated in \autoref{tab:table2}. We found that the participants obtained unsatisfactory accuracies on Q16 (accuracy: $\mu$ = 0.45) and Q25 (accuracy: $\mu$ = 0.60), as shown in \autoref{tab:table1}. Q16 and Q25 asked the participants to judge the recent transaction trends of a certain address in the address page. The participants were allowed to freely select addresses to answer the two questions. Thus, some participants encountered addresses with ambiguous trends or no recent transactions.

\textit{\textbf{(2) Result analysis by exploring mode.}}

As shown in \autoref{tab:table3}, the query exploring mode is slightly better than the hierarchical exploring mode in terms of all four metrics. This result was highly interesting and worthy of an in-depth further study. First, this result did not indicate that the hierarchical exploring mode can be disregarded. The majority of NUsers confirmed that the hierarchical exploring mode helped them understand the four-level structure of the transaction data. Many EUsers said that the two exploring modes complemented each other during data viewing. Second, we had two preliminary interpretations for this result. (1) The questions regarding the query exploring mode provided the specific ID of a block, transaction, or address. Thus, the participants had a clear target and strong motivation. (2) The visual stimulations provided by the visualizations and received by the participants may gradually decline during the hierarchical browsing of the four levels of web pages.
\begin{table}[!h]
	\scriptsize
	\renewcommand\tabcolsep{8.5pt}  
	\renewcommand\arraystretch{1.2} 
	\centering
	\vspace{-0.15cm}  
	\setlength{\belowcaptionskip}{-0.5cm}   
	\caption{Objective questionnaire results in terms of mean accuracy, mean time, mean easiness score, and mean confidence score by exploring mode (H: hierarchical exploring mode and Q: query exploring mode).}
	\label{tab:table3}
	\addvbuffer[0pt -5pt]{
	\begin{tabular}{m{2cm}<{\centering}m{1cm}<{\centering}m{0.5cm}<{\centering}m{0.8cm}<{\centering}m{1cm}<{\centering}}
\hline
Exploring mode &	Accuracy &	Time &	Easiness&	\textls[-50]{Confidence}\\
\hline
H &	0.79 &	65.19 &	3.69 &	3.70\\
Q &	0.81 &	47.05 &	3.87 &	3.90\\
\hline
\end{tabular}
}
\end{table}

\textit{\textbf{(3) Result analysis by user type}}

In general, the 22 EUsers were more familiar with relevant knowledge and more skilled with operating SilkViser than the 23 NUsers. The EUsers should exhibit evident advantages in completing all the objective questions. The results in \autoref{tab:table1} indicated that significant differences between the two types of users were found in many questions in terms of time, easiness, and confidence. For example, Q2 inquired which information field in a block connects with the previous block. 8 EUsers said that they knew the correct answer even without looking at the options provided by Q2. Therefore, the EUsers significantly outperformed the NUsers in terms of time, easiness, and confidence. Q10 asked the participants to find the transaction with a high transaction fee in a block. The EUsers were more proficient than the NUsers in using brushing and sorting interactions to obtain the answer, resulting in the EUsers' higher performance in terms of time, easiness, and confidence. However, no significant differences existed between the EUsers and the NUsers in terms of accuracy in all questions, indicating that our visual design can compensate for the inexperience of the NUsers to a certain extent in completing the questions.
 \begin{figure}[htb]
	\centering
	\vspace{-0.2cm}  
	\setlength{\abovecaptionskip}{-0.03cm}   
	\setlength{\belowcaptionskip}{-0.4cm}   
	\includegraphics[width=0.98\columnwidth]{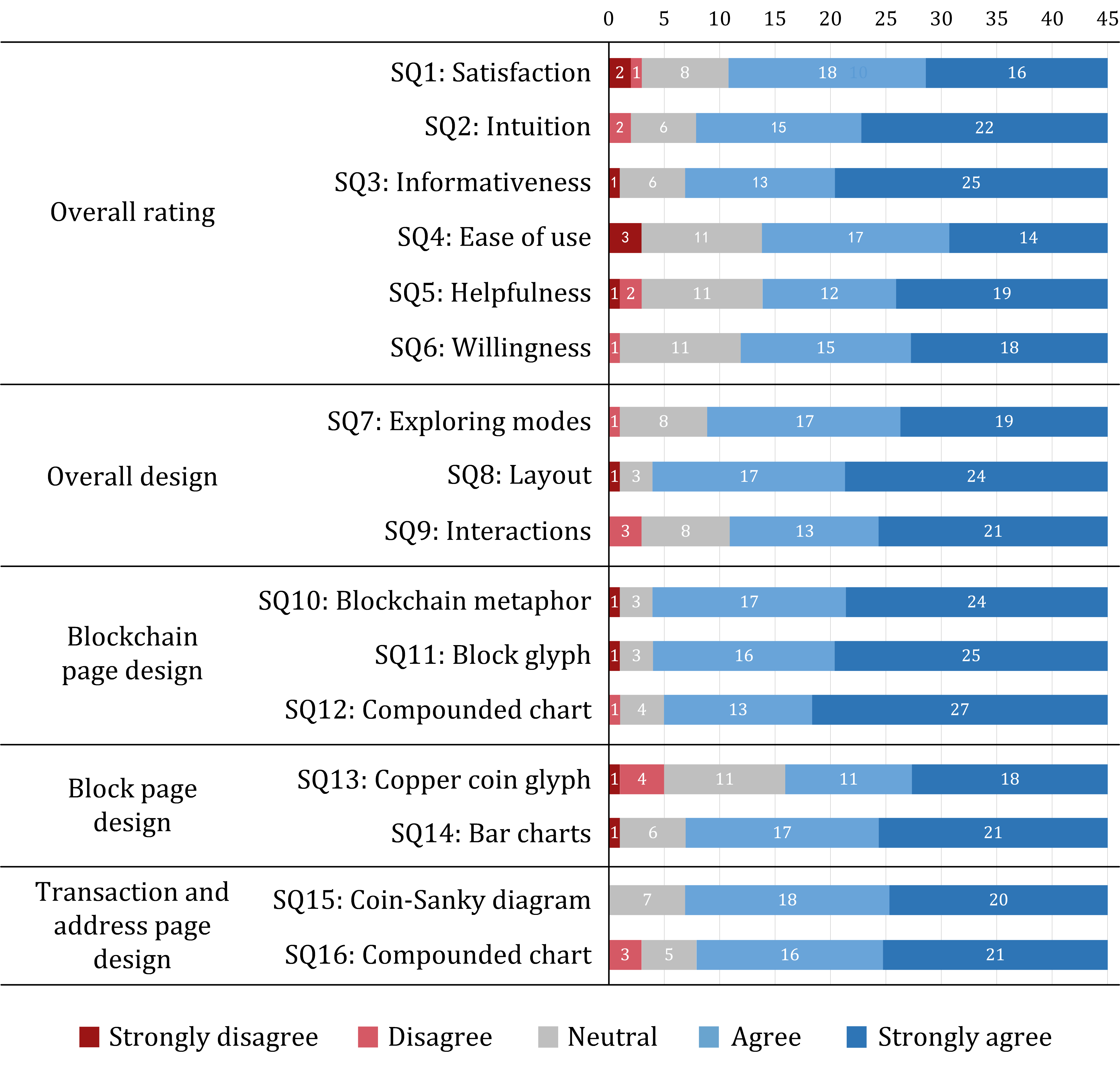}
	\caption{Stacked bar chart of the subjective ratings. Each participant answered 16 SQs in the subjective questionnaire with a five-point Likert scale at the end of the formal experiment.}
	\label{figure7}
\end{figure}

\subsubsection{Analysis of subjective questionnaire results}
\autoref{figure7} shows the ratings of all the participants on the 16 SQs. For SQ1--SQ6, the participants' overall feelings toward SilkViser were relatively good. In particular, the rating result of SQ6 encouraged us. All the EUsers confirmed that they were willing to use SilkViser instead of SilkViewer. They commended that SilkViser had fully covered SilkViewer's functions and satisfied their requirement for advanced information acquisition. Most of the NUsers stated that using SilkViser increased their confidence in participating in actual Silubium transactions. For SQ7--SQ9, the participants were generally satisfied with the exploring mode design, layout design, and interaction design. The results of SQ10--SQ16 indicated that the participants were generally satisfied with all the web pages. Among them, the blockchain page received the highest ratings. Nearly all the participants highly praised this page. Most of the participants liked the design of the coin-Sankey diagram in the transaction page. The block page design obtained a relatively low score. Some of the participants expressed that the block page caused the highest learning cost among all the designs.
\section{Discussion}

In this section, we discuss the limitations of this work and suggest directions for further work.

SilkViser is implemented via HTML5 and supports a variety of terminals. However, the interface has not been optimized well to fit all web browsers and various display sizes of mobile terminals. At present, users can only achieve the best usage experience in desktop browsers. Silubium originates from China. Thus, we add Chinese elements into SilkViser, such as paper ledgers and copper coins. However, we are unsure if clients from other countries will accept these Chinese elements. The visualization designs of SilkViser have some defects. First, the visualization design of the block page should be further optimized to reduce users’ learning cost; Second, SilkViser cannot support the presentations of large-scale blocks and transactions; At last, the compounded line and bar charts in the blockchain and address pages may produce a potentially confusing understanding of the values. 

We did not compare SilkViser with SilkViewer in the user study based on two considerations. First, SilkViser maintains most of the textual and tabular information presentation forms of SilkViewer and its functions. Second, we focused on evaluating the visualization designs of SilkViser, but SilkViewer has only minimal visualizations. Therefore, eliminating a comparison between SilkViser and SilkViewer can simplify the formal experiment and shorten the time consumed by the volunteers. Evaluations in other forms or the long term should still be conducted among a wide range of users.

SilkViser is currently not a visual analytics application, lacking analysis capability. We have two considerations on this issue. First, a large proportion of cryptocurrency users have not an IT or data analysis background. Blockchain explorers have to meet the requirements of information acquisition of most users. Second, a small part of users is data analysts or administrators who are interested in gaining in-depth insights from the data. We should provide them with other advanced visual analytical tools to dig into the data and find out complex data patterns.

Our future work can be conducted in three directions. (1) Improve visual designs, such as solving the learning cost issue of the block page and the scalability issue of displaying blocks and transactions. (2) Attempt to extend SilkViser to viewing the transaction data of other cryptocurrencies. We have considered this point in an early stage. Our data abstraction and requirement analysis focus on the possibility of sharing our designs among various cryptocurrencies. SilkViser can theoretically be used to explore the transaction data of other cryptocurrencies with only minor modifications. But we are unsure if the visualization designs of SilkViser can be used for visualizing other data forms, such as fiat currency transaction data. (3) Design advanced visual analytical tools for in-depth exploration of complex transaction patterns.

\section{Conclusion}
This study presents a web tool for viewing blockchain-based cryptocurrency transaction data, called SilkViser. SilkViser adopts a number of appropriating designs with visual data presentations. The visualizations can help NUsers understand important concepts and processes in cryptocurrency transaction systems. The visualizations can also satisfy EUsers' requirements for advanced information acquisition and improve their efficiency of viewing transaction data. We hope that SilkViser will inspire other cryptocurrency providers to create novel transaction data viewing tools or services. Moreover, SilkViser should increase the confidence of researchers, designers, and developers in the cryptocurrency community to apply visualization and visual analysis technologies to a wide range of scenarios in blockchain data presentation and analysis.
\acknowledgments{
	We wish to thank the reviewers for their thoughtful comments, the volunteers for their active involvements, and Xin Huang from Layer Vis Lab at Qi An Xin Technology Group for the fruitful discussions. The work is supported in part by the National Key Research and Development Program of China (No. 2018YFB1700403), the National Natural Science Foundation of China (No. 61872388, 61672538), and the Research Platform Open Project of Chongqing Technology and Business University (No. KFJJ2019079). SilkViser at Github: https://github.com/csuvis/BlockchainVis/.
}
\bibliographystyle{abbrv-doi}
\bibliography{myreference-doi}

\begin{thebibliography}{10}

\bibitem{A7}
\url{https://github.com/csuvis/BlockchainVis/}.
\newblock Accessed: 18-Aug-2019.

\bibitem{C4}
Bitbonkers.
\newblock \url{https://bitbonkers.com/}.
\newblock Accessed: 2-Sep-2018.

\bibitem{C3}
Bitcoin transaction visualization.
\newblock \url{http://bitcoin.interaqt.nl/}.
\newblock Accessed: 23-Aug-2018.

\bibitem{C5}
Bitcoincity.info - the road to the blockchain.
\newblock \url{http://bitcoincity.info/}.
\newblock Accessed: 23-May-2019.

\bibitem{C13}
Bitstamp btc/usd charts - bitcoinwisdom.
\newblock \url{https://bitcoinwisdom.com/markets/bitstamp/btcusd}.
\newblock Accessed: 23-Septemper-2019.

\bibitem{A4}
Blockchain explorer--search the blockchain | btc | eth | bch.
\newblock \url{https://www.blockchain.com/explorer}.
\newblock Accessed: 10-November-2018.

\bibitem{A1}
A brief history of bitcoin.
\newblock \url{http://dataconomy.com/2017/07/history-of-bitcoin/}.
\newblock Accessed: 4-April-2019.

\bibitem{C10}
Cryptocurrency market capitalizations | coinmarketcap.
\newblock \url{https://coinmarketcap.com/}.
\newblock Accessed: 23-July-2019.

\bibitem{C11}
Cryptocurrency prices heatmap, market cap, charts widget - coin360.
\newblock \url{https://coin360.com/}.
\newblock Accessed: 3-July-2019.

\bibitem{C7}
Global bitcoin nodes distribution--bitnodes.
\newblock Accessed: 15-May-2019.

\bibitem{A2}
How many cryptocurrencies are there? blocklr.
\newblock \url{http://blocklr.com/guides/how-many-cryptocurrencies-are-there/}.
\newblock Accessed: 20-August-2018.

\bibitem{A3}
How many people use bitcoin in 2019? - bitcoin market journal.
\newblock
  \url{https://www.bitcoinmarketjournal.com/how-many-people-use-bitcoin/}.
\newblock Accessed: 20-Septemper-2018.

\bibitem{A6}
Litecoin block explorer | blockcypher.
\newblock \url{https://live.blockcypher.com/ltc/}.
\newblock Accessed: 8-December-2018.

\bibitem{C8}
Realtime bitcoin globe.
\newblock \url{https://blocks.wizb.it}.
\newblock Accessed: 28-July-2019.

\bibitem{C14}
Tradeblock.
\newblock \url{https://tradeblock.com/markets/bfnx/xbt-usd/5m/}.
\newblock Accessed: 25-Septemper-2019.

\bibitem{A5}
Zcash (zec) block explorer--minergate.
\newblock \url{https://minergate.com/blockchain/zec/blocks}.
\newblock Accessed: 28-Septemper-2018.

\bibitem{B10}
S.~Athey, I.~Parashkevov, V.~Sarukkai, and J.~Xia.
\newblock Bitcoin pricing, adoption, and usage: Theory and evidence.
\newblock {\em Research Papers}, 2016.

\bibitem{C15}
G.~D. {Battista}, V.~D. {Donato}, M.~{Patrignani}, M.~{Pizzonia}, V.~{Roselli},
  and R.~{Tamassia}.
\newblock Bitconeview: visualization of flows in the bitcoin transaction graph.
\newblock In {\em 2015 IEEE Symposium on Visualization for Cyber Security
  (VizSec)}, pp. 1--8. IEEE, 2015. doi: {{%
10\hspace{.1pt}\discretionary{.}{%
}{.}\hspace{.4pt}1109\discretionary{/}{%
}{/}VIZSEC\hspace{.1pt}\discretionary{.}{%
}{.}\hspace{.4pt}2015\hspace{.1pt}\discretionary{.}{%
}{.}\hspace{.4pt}7312773}}


\bibitem{F2}
M.~Behrisch, D.~Streeb, F.~Stoffel, D.~Seebacher, B.~Matejek, S.~Weber,
  S.~Mittelst\"{a}dt, H.~Pfister, and D.~Keim.
\newblock Commercial visual analytics systems-advances in the big data
  analytics field.
\newblock {\em IEEE Transactions on Visualization and Computer Graphics},
  25(10):3011--3031, 2018. doi: {{%
10\hspace{.1pt}\discretionary{.}{%
}{.}\hspace{.4pt}1109\discretionary{/}{%
}{/}TVCG\hspace{.1pt}\discretionary{.}{%
}{.}\hspace{.4pt}2018\hspace{.1pt}\discretionary{.}{%
}{.}\hspace{.4pt}2859973}}


\bibitem{C19}
S.~Bistarelli and F.~Santini.
\newblock Go with the -bitcoin- flow, with visual analytics.
\newblock In {\em the 12th International Conference on Availability,
  Reliability and Security}, pp. 1--6. ACM, 2017. doi: {{%
10\hspace{.1pt}\discretionary{.}{%
}{.}\hspace{.4pt}1145\discretionary{/}{%
}{/}3098954\hspace{.1pt}\discretionary{.}{%
}{.}\hspace{.4pt}3098972}}


\bibitem{F6}
R.~Blinder, O.~Biller, A.~Even, O.~Sofer, N.~Tractinsky, J.~Lanir, and P.~Bak.
\newblock Comparative evaluation of node-link and sankey diagrams for the cyber
  security domain.
\newblock In {\em IFIP Conference on Human-Computer Interaction}, pp. 497--518.
  Springer, 2019. doi: {{%
10\hspace{.1pt}\discretionary{.}{%
}{.}\hspace{.4pt}1007\discretionary{/}{%
}{/}978\discretionary{%
}{-}{-}3\discretionary{%
}{-}{-}030\discretionary{%
}{-}{-}29381\discretionary{%
}{-}{-}9\_31}}


\bibitem{B14}
J.~Bohr and M.~Bashir.
\newblock Who uses bitcoin? an exploration of the bitcoin community.
\newblock In {\em 2014 Twelfth Annual International Conference on Privacy,
  Security and Trust}, pp. 94--101. IEEE, 2014. doi: {{%
10\hspace{.1pt}\discretionary{.}{%
}{.}\hspace{.4pt}1109\discretionary{/}{%
}{/}PST\hspace{.1pt}\discretionary{.}{%
}{.}\hspace{.4pt}2014\hspace{.1pt}\discretionary{.}{%
}{.}\hspace{.4pt}6890928}}


\bibitem{F11}
B.~Chongke, Y.~Lu, D.~Yulin, and S.~Yun.
\newblock A survey on visualization of tensor field.
\newblock {\em Journal of Visualization}, 22(3):641--660, 2019. doi: {{%
10\hspace{.1pt}\discretionary{.}{%
}{.}\hspace{.4pt}1007\discretionary{/}{%
}{/}s12650\discretionary{%
}{-}{-}019\discretionary{%
}{-}{-}00555\discretionary{%
}{-}{-}8}}


\bibitem{C20}
A.~Elbahrawy, L.~Alessandretti, A.~Kandler, R.~Pastor-Satorras, and
  A.~Baronchelli.
\newblock Evolutionary dynamics of the cryptocurrency market.
\newblock {\em Royal Society Open Science}, 4(11):170623--170632, 2017. doi:
  {{%
10\hspace{.1pt}\discretionary{.}{%
}{.}\hspace{.4pt}1098\discretionary{/}{%
}{/}rsos\hspace{.1pt}\discretionary{.}{%
}{.}\hspace{.4pt}170623}}


\bibitem{F1}
Z.~Fangfang, L.~Xiaoru, L.~Chang, Z.~Ying, X.~Panpan, L.~Ren, X.~Tingmin, and
  R.~Lei.
\newblock A survey of visualization for smart manufacturing.
\newblock {\em Journal of Visualization}, 22(2):419--435, 2019. doi: {{%
10\hspace{.1pt}\discretionary{.}{%
}{.}\hspace{.4pt}1007\discretionary{/}{%
}{/}s12650\discretionary{%
}{-}{-}018\discretionary{%
}{-}{-}0530\discretionary{%
}{-}{-}2}}


\bibitem{F3}
Z.~Fangfang, L.~Xiaoru, L.~Xiaobo, Z.~Ying, C.~Yi, C.~Ning, and G.~Weihua.
\newblock Visually enhanced situation awareness for complex manufacturing
  facility monitoring in smart factories.
\newblock {\em Journal of Visual Languages \& Computing}, 44(2):58--69, 2017.
  doi: {{%
10\hspace{.1pt}\discretionary{.}{%
}{.}\hspace{.4pt}1016\discretionary{/}{%
}{/}j\hspace{.1pt}\discretionary{.}{%
}{.}\hspace{.4pt}jvlc\hspace{.1pt}\discretionary{.}{%
}{.}\hspace{.4pt}2017\hspace{.1pt}\discretionary{.}{%
}{.}\hspace{.4pt}11\hspace{.1pt}\discretionary{.}{%
}{.}\hspace{.4pt}004}}


\bibitem{D6}
M.~Fleder, M.~S. Kester, and S.~Pillai.
\newblock Bitcoin transaction graph analysis.
\newblock {\em arXiv preprint arXiv:1502.01657}, abs/1502.01657.

\bibitem{B3}
A.~E. Gencer, S.~Basu, I.~Eyal, R.~Van~Renesse, and E.~G. Sirer.
\newblock Decentralization in bitcoin and ethereum networks.
\newblock In {\em the 22nd International Conference on Financial Cryptography
  and Data Security}, pp. 439--457. Springer, 2018. doi: {{%
10\hspace{.1pt}\discretionary{.}{%
}{.}\hspace{.4pt}1007\discretionary{/}{%
}{/}978\discretionary{%
}{-}{-}3\discretionary{%
}{-}{-}662\discretionary{%
}{-}{-}58387\discretionary{%
}{-}{-}6\_24}}


\bibitem{B8}
Gronwald and Marc.
\newblock The economics of bitcoins - market characteristics and price jumps.
\newblock {\em University of Aberdeen; CESifo(Center for Economic Studies and
  Ifo Institute)}, 2014.

\bibitem{D4}
P.~Isenberg, C.~Kinkeldey, and J.-D. Fekete.
\newblock Exploring entity behavior on the bitcoin blockchain.
\newblock In {\em IEEE Conference on Visualization, VIS 2017, Phoenix, Arizona
  USA}, pp. 1--2. IEEE, 2017.

\bibitem{F14}
X.~Jiazhi, Y.~Fenjin, C.~Wei, W.~Yusi, C.~Weifeng, M.~Yuxin, and A.~K. Tung.
\newblock Ldsscanner: Exploratory analysis of low-dimensional structures in
  high-dimensional datasets.
\newblock {\em IEEE Transactions on Visualization and Computer Graphics},
  24(1):236--245, 2018. doi: {{%
10\hspace{.1pt}\discretionary{.}{%
}{.}\hspace{.4pt}1109\discretionary{/}{%
}{/}tvcg\hspace{.1pt}\discretionary{.}{%
}{.}\hspace{.4pt}2017\hspace{.1pt}\discretionary{.}{%
}{.}\hspace{.4pt}2744098}}


\bibitem{C22}
X.~Jiazhi, Z.~Yuhong, Y.~Hui, W.~Ying, J.~Guang, Z.~Ying, X.~Cong, K.~Xiaoyan,
  L.~Shenghui, and W.~Weiping.
\newblock Supoolvisor: a visual analytics system for mining pool surveillance.
\newblock {\em Frontiers of Information Technology \& Electronic Engineering},
  21(4):507–523, 2020. doi: {{%
10\hspace{.1pt}\discretionary{.}{%
}{.}\hspace{.4pt}1631\discretionary{/}{%
}{/}FITEE\hspace{.1pt}\discretionary{.}{%
}{.}\hspace{.4pt}1900532}}


\bibitem{B13}
Y.~B. Kim, J.~Kim, W.~Kim, J.~Im, T.~Kim, S.~Kang, and C.-H. Kim.
\newblock Predicting fluctuations in cryptocurrency transactions based on user
  comments and replies.
\newblock {\em PLOS ONE}, 11:e0161197, 2016. doi: {{%
10\hspace{.1pt}\discretionary{.}{%
}{.}\hspace{.4pt}1371\discretionary{/}{%
}{/}journal\hspace{.1pt}\discretionary{.}{%
}{.}\hspace{.4pt}pone\hspace{.1pt}\discretionary{.}{%
}{.}\hspace{.4pt}0161197}}


\bibitem{D5}
C.~Kinkeldey, J.~Fekete, and P.~Isenberg.
\newblock Bitconduite: Visualizing and analyzing activity on the bitcoin
  network.
\newblock In {\em \textls[-25]{Eurographics Conference on Visualization,
  EuroVis 2017, Barcelona, Spain}}, pp. 25--27. Eurographics Association, 2017.
  doi: {{%
10\hspace{.1pt}\discretionary{.}{%
}{.}\hspace{.4pt}2312\discretionary{/}{%
}{/}eurp\hspace{.1pt}\discretionary{.}{%
}{.}\hspace{.4pt}20171160}}


\bibitem{B7}
M.~Kiran and M.~Stanett.
\newblock Bitcoin risk analysis.
\newblock {\em NEMODE Policy Paper}, abs/1502.01657, 2015.

\bibitem{D3}
P.~Koshy, D.~Koshy, and P.~McDaniel.
\newblock An analysis of anonymity in bitcoin using p2p network traffic.
\newblock In {\em \textls[-25]{International Conference on Financial
  Cryptography and Data Security}}, pp. 469--485. Springer, 2014. doi: {{%
10\hspace{.1pt}\discretionary{.}{%
}{.}\hspace{.4pt}1007\discretionary{/}{%
}{/}978\discretionary{%
}{-}{-}3\discretionary{%
}{-}{-}662\discretionary{%
}{-}{-}45472\discretionary{%
}{-}{-}5\_30}}


\bibitem{D2}
I.-K. Lim, Y.-H. Kim, J.-G. Lee, J.-P. Lee, H.~Nam-Gung, and J.-K. Lee.
\newblock The analysis and countermeasures on security breach of bitcoin.
\newblock In {\em International Conference on Computational Science and Its
  Applications}, pp. 720--732. Springer, 2014. doi: {{%
10\hspace{.1pt}\discretionary{.}{%
}{.}\hspace{.4pt}1007\discretionary{/}{%
}{/}978\discretionary{%
}{-}{-}3\discretionary{%
}{-}{-}319\discretionary{%
}{-}{-}09147\discretionary{%
}{-}{-}1\_52}}


\bibitem{F9}
M.~Liu, J.~Shi, Z.~Li, C.~Li, J.~Zhu, and S.~Liu.
\newblock Towards better analysis of deep convolutional neural networks.
\newblock {\em IEEE Transactions on Visualization and Computer Graphics},
  23(1):91--100, 2017. doi: {{%
10\hspace{.1pt}\discretionary{.}{%
}{.}\hspace{.4pt}1109\discretionary{/}{%
}{/}TVCG\hspace{.1pt}\discretionary{.}{%
}{.}\hspace{.4pt}2016\hspace{.1pt}\discretionary{.}{%
}{.}\hspace{.4pt}2598831}}


\bibitem{C6}
L.~Matthias and F.~Benjamin.
\newblock Analyzing the bitcoin network: The first four years.
\newblock {\em Future Internet}, 8(1):1--7, 2016. doi: {{%
10\hspace{.1pt}\discretionary{.}{%
}{.}\hspace{.4pt}3390\discretionary{/}{%
}{/}fi8010007}}


\bibitem{C16}
D.~McGinn, D.~Birch, D.~Akroyd, M.~Molina-Solana, Y.~Guo, and W.~Knottenbelt.
\newblock Visualizing dynamic bitcoin transaction patterns.
\newblock {\em Big Data}, 4:109--119, 2016. doi: {{%
10\hspace{.1pt}\discretionary{.}{%
}{.}\hspace{.4pt}1089\discretionary{/}{%
}{/}big\hspace{.1pt}\discretionary{.}{%
}{.}\hspace{.4pt}2015\hspace{.1pt}\discretionary{.}{%
}{.}\hspace{.4pt}0056}}


\bibitem{F18}
H.~Mei, H.~Guan, C.~Xin, X.~Wen, and W.~Chen.
\newblock Datav: Data visualization on large high-resolution displays.
\newblock {\em Visual Informatics}, 4(3):12--23, 2020. doi: {{%
10\hspace{.1pt}\discretionary{.}{%
}{.}\hspace{.4pt}1016\discretionary{/}{%
}{/}j\hspace{.1pt}\discretionary{.}{%
}{.}\hspace{.4pt}visinf\hspace{.1pt}\discretionary{.}{%
}{.}\hspace{.4pt}2020\hspace{.1pt}\discretionary{.}{%
}{.}\hspace{.4pt}07\hspace{.1pt}\discretionary{.}{%
}{.}\hspace{.4pt}001}}


\bibitem{B6}
S.~Meiklejohn, M.~Pomarole, G.~Jordan, K.~Levchenko, D.~McCoy, G.~M. Voelker,
  and S.~Savage.
\newblock A fistful of bitcoins: Characterizing payments among men with no
  names.
\newblock In {\em the 2013 conference on Internet measurement conference}, pp.
  127--140. ACM, 2013. doi: {{%
10\hspace{.1pt}\discretionary{.}{%
}{.}\hspace{.4pt}1145\discretionary{/}{%
}{/}2504730\hspace{.1pt}\discretionary{.}{%
}{.}\hspace{.4pt}2504747}}


\bibitem{B12}
T.~Moore and N.~Christin.
\newblock Beware the middleman: Empirical analysis of bitcoin-exchange risk.
\newblock In {\em the 17th International Conference on Financial Cryptography
  and Data Security, Okinawa, Japan}, pp. 25--33. Springer, 2013. doi: {{%
10\hspace{.1pt}\discretionary{.}{%
}{.}\hspace{.4pt}1007\discretionary{/}{%
}{/}978\discretionary{%
}{-}{-}3\discretionary{%
}{-}{-}642\discretionary{%
}{-}{-}39884\discretionary{%
}{-}{-}1\_3}}


\bibitem{B4}
M.~M\"{o}ser, R.~Bohme, and D.~Breuker.
\newblock An inquiry into money laundering tools in the bitcoin ecosystem.
\newblock In {\em the 2013 APWG eCrime Researchers Summit}, pp. 1--14. IEEE,
  2013. doi: {{%
10\hspace{.1pt}\discretionary{.}{%
}{.}\hspace{.4pt}1109\discretionary{/}{%
}{/}eCRS\hspace{.1pt}\discretionary{.}{%
}{.}\hspace{.4pt}2013\hspace{.1pt}\discretionary{.}{%
}{.}\hspace{.4pt}6805780}}


\bibitem{F19}
T.~Natkamon, H.~Nicolas, F.~Jean-Daniel, and I.~Petra.
\newblock Visualization of blockchain data: A systematic review.
\newblock {\em IEEE Transactions on Visualization and Computer Graphics}, 2020.
  doi: {{%
10\hspace{.1pt}\discretionary{.}{%
}{.}\hspace{.4pt}1109\discretionary{/}{%
}{/}TVCG\hspace{.1pt}\discretionary{.}{%
}{.}\hspace{.4pt}2019\hspace{.1pt}\discretionary{.}{%
}{.}\hspace{.4pt}2963018}}


\bibitem{C17}
T.~Pham and S.~Lee.
\newblock Anomaly detection in the bitcoin system - {A} network perspective.
\newblock {\em arXiv preprint arXiv:1611.03942}, abs/1611.03942, 2016.

\bibitem{B2}
F.~Reid and M.~Harrigan.
\newblock An analysis of anonymity in the bitcoin system.
\newblock {\em Security and Privacy in Social Networks}, 3:197--223, 2011. doi:
  {{%
10\hspace{.1pt}\discretionary{.}{%
}{.}\hspace{.4pt}1109\discretionary{/}{%
}{/}PASSAT\discretionary{/}{%
}{/}SocialCom\hspace{.1pt}\discretionary{.}{%
}{.}\hspace{.4pt}2011\hspace{.1pt}\discretionary{.}{%
}{.}\hspace{.4pt}79}}


\bibitem{B5}
D.~Ron and A.~Shamir.
\newblock Quantitative analysis of the full bitcoin transaction graph.
\newblock In {\em the 17th International Conference on Financial Cryptography
  and Data Security, Okinawa, Japan}, pp. 6--24. Springer, 2012. doi: {{%
10\hspace{.1pt}\discretionary{.}{%
}{.}\hspace{.4pt}1007\discretionary{/}{%
}{/}978\discretionary{%
}{-}{-}3\discretionary{%
}{-}{-}642\discretionary{%
}{-}{-}39884\discretionary{%
}{-}{-}1\_2}}


\bibitem{F7}
Schmidt and Mario.
\newblock The sankey diagram in energy and material flow management - part ii:
  Methodology and current applications.
\newblock {\em Journal of Industrial Ecology}, 12:173--185, 2008. doi: {{%
10\hspace{.1pt}\discretionary{.}{%
}{.}\hspace{.4pt}1111\discretionary{/}{%
}{/}j\hspace{.1pt}\discretionary{.}{%
}{.}\hspace{.4pt}1530\discretionary{%
}{-}{-}9290\hspace{.1pt}\discretionary{.}{%
}{.}\hspace{.4pt}2008\hspace{.1pt}\discretionary{.}{%
}{.}\hspace{.4pt}00004\hspace{.1pt}\discretionary{.}{%
}{.}\hspace{.4pt}x}}


\bibitem{C18}
Sudarshan and Chawathe.
\newblock Monitoring blockchains with self-organizing maps.
\newblock In {\em the 17th IEEE International Conference On Trust, Security And
  Privacy In Computing And Communications}, pp. 1870--1875. IEEE, 2018. doi:
  {{%
10\hspace{.1pt}\discretionary{.}{%
}{.}\hspace{.4pt}1109\discretionary{/}{%
}{/}TrustCom\discretionary{/}{%
}{/}BigDataSE\hspace{.1pt}\discretionary{.}{%
}{.}\hspace{.4pt}2018\hspace{.1pt}\discretionary{.}{%
}{.}\hspace{.4pt}00283}}


\bibitem{B1}
F.~Tschorsch and B.~Scheuermann.
\newblock Bitcoin and beyond: A technical survey on decentralized digital
  currencies.
\newblock {\em IEEE Communications Surveys \& Tutorials}, 18(3):2084--2123,
  2016. doi: {{%
10\hspace{.1pt}\discretionary{.}{%
}{.}\hspace{.4pt}1109\discretionary{/}{%
}{/}COMST\hspace{.1pt}\discretionary{.}{%
}{.}\hspace{.4pt}2016\hspace{.1pt}\discretionary{.}{%
}{.}\hspace{.4pt}2535718}}


\bibitem{D1}
M.~Vasek and T.~Moore.
\newblock There's no free lunch, even using bitcoin: Tracking the popularity
  and profits of virtual currency scams.
\newblock In {\em the 19th International conference on Financial Cryptography
  and Data Security}, pp. 44--61. Springer, 2015. doi: {{%
10\hspace{.1pt}\discretionary{.}{%
}{.}\hspace{.4pt}1007\discretionary{/}{%
}{/}978\discretionary{%
}{-}{-}3\discretionary{%
}{-}{-}662\discretionary{%
}{-}{-}47854\discretionary{%
}{-}{-}7\_4}}


\bibitem{F17}
X.~Wang, W.~Chen, J.-K. Chou, C.~Bryan, H.~Guan, W.~Chen, R.~Pan, and K.-L. Ma.
\newblock Graphprotector: A visual interface for employing and assessing
  multiple privacy preserving graph algorithms.
\newblock {\em IEEE transactions on visualization and computer graphics},
  25(1):193--203, 2019. doi: {{%
10\hspace{.1pt}\discretionary{.}{%
}{.}\hspace{.4pt}1109\discretionary{/}{%
}{/}TVCG\hspace{.1pt}\discretionary{.}{%
}{.}\hspace{.4pt}2018\hspace{.1pt}\discretionary{.}{%
}{.}\hspace{.4pt}2865021}}


\bibitem{F15}
S.~Yang, B.~Chris, B.~Sridatt, Z.~Ying, Z.~Yaoxue, and M.~Kwan-Liu.
\newblock Meetingvis: Visual narratives to assist in recalling meeting context
  and content.
\newblock {\em IEEE Transactions on Visualization and Computer Graphics},
  24(6):1918--1929, 2018. doi: {{%
10\hspace{.1pt}\discretionary{.}{%
}{.}\hspace{.4pt}1109\discretionary{/}{%
}{/}TVCG\hspace{.1pt}\discretionary{.}{%
}{.}\hspace{.4pt}2018\hspace{.1pt}\discretionary{.}{%
}{.}\hspace{.4pt}2816203}}


\bibitem{F12}
W.~Yating, M.~Honghui, Z.~Ying, Z.~Shuyue, L.~Bingru, J.~Haojin, and C.~Wei.
\newblock Evaluating perceptual bias during geometric scaling of scatterplots.
\newblock {\em IEEE Transactions on Visualization and Computer Graphics},
  26(1):100--111, 2020. doi: {{%
10\hspace{.1pt}\discretionary{.}{%
}{.}\hspace{.4pt}1109\discretionary{/}{%
}{/}TVCG\hspace{.1pt}\discretionary{.}{%
}{.}\hspace{.4pt}2019\hspace{.1pt}\discretionary{.}{%
}{.}\hspace{.4pt}2934208}}


\bibitem{F8}
C.~Yi, G.~Zeli, Z.~Rong, D.~Xiaomin, and W.~Yunhai.
\newblock A survey on visualization approaches for exploring association
  relationships in graph data.
\newblock {\em Journal of Visualization}, 22(3):625--639, 2019. doi: {{%
10\hspace{.1pt}\discretionary{.}{%
}{.}\hspace{.4pt}1007\discretionary{/}{%
}{/}s12650\discretionary{%
}{-}{-}019\discretionary{%
}{-}{-}00551\discretionary{%
}{-}{-}y}}


\bibitem{F5}
Z.~Ying, L.~Feng, C.~Minghui, W.~Yingchao, X.~Jiazhi, Z.~Fangfang, W.~Yunhai,
  C.~Yi, and C.~Wei.
\newblock Evaluating multi-dimensional visualizations for understanding fuzzy
  clusters.
\newblock {\em IEEE Transactions on Visualization and Computer Graphics},
  25(1):12--21, 2019. doi: {{%
10\hspace{.1pt}\discretionary{.}{%
}{.}\hspace{.4pt}1109\discretionary{/}{%
}{/}TVCG\hspace{.1pt}\discretionary{.}{%
}{.}\hspace{.4pt}2018\hspace{.1pt}\discretionary{.}{%
}{.}\hspace{.4pt}2865020}}


\bibitem{F10}
Z.~Ying, W.~Lei, L.~Shijie, Z.~Fangfang, L.~Xiaoru, L.~Qiang, and R.~Lei.
\newblock A visual analysis approach for understanding durability test data of
  automotive products.
\newblock {\em ACM Transactions on Intelligent Systems and Technology},
  10(6):70--93, 2019. doi: {{%
10\hspace{.1pt}\discretionary{.}{%
}{.}\hspace{.4pt}1145\discretionary{/}{%
}{/}3345640}}


\bibitem{F4}
Z.~Ying, L.~Xiaobo, L.~Xiaoru, W.~Hairong, K.~Xiaoyan, Z.~Fangfang, W.~Jinsong,
  C.~Yi, and C.~Wei.
\newblock Visual analytics for electromagnetic situation awareness in radio
  monitoring and management.
\newblock {\em IEEE Transactions on Visualization and Computer Graphics},
  26(1):590--600, 2020. doi: {{%
10\hspace{.1pt}\discretionary{.}{%
}{.}\hspace{.4pt}1109\discretionary{/}{%
}{/}TVCG\hspace{.1pt}\discretionary{.}{%
}{.}\hspace{.4pt}2019\hspace{.1pt}\discretionary{.}{%
}{.}\hspace{.4pt}2934655}}


\bibitem{F16}
J.~Yuan, C.~Chen, W.~Yang, M.~Liu, J.~Xia, and S.~Liu.
\newblock A survey of visual analytics techniques for machine learning.
\newblock {\em Computational Visual Media}, 7(1):1--31, 2021.

\bibitem{C21}
X.~Yue, X.~Shu, X.~Zhu, X.~Du, Z.~Yu, D.~Papadopoulos, and S.~Liu.
\newblock Bitextract: Interactive visualization for extracting bitcoin exchange
  intelligence.
\newblock {\em IEEE Transactions on Visualization and Computer Graphics},
  25(1):162--171, 2018. doi: {{%
10\hspace{.1pt}\discretionary{.}{%
}{.}\hspace{.4pt}1109\discretionary{/}{%
}{/}TVCG\hspace{.1pt}\discretionary{.}{%
}{.}\hspace{.4pt}2018\hspace{.1pt}\discretionary{.}{%
}{.}\hspace{.4pt}2864814}}


\bibitem{F13}
Z.~Zhiguang, M.~Linhao, T.~Cheng, Z.~Ying, G.~Zhiyong, H.~Miaoxin, and C.~Wei.
\newblock Visual abstraction of large scale geospatial origin-destination
  movement data.
\newblock {\em IEEE Transactions on Visualization and Computer Graphics},
  25(1):43--53, 2018. doi: {{%
10\hspace{.1pt}\discretionary{.}{%
}{.}\hspace{.4pt}1109\discretionary{/}{%
}{/}TVCG\hspace{.1pt}\discretionary{.}{%
}{.}\hspace{.4pt}2018\hspace{.1pt}\discretionary{.}{%
}{.}\hspace{.4pt}2864503}}


\end{thebibliography}

\end{document}